\documentclass[longauth]{aaEC}

\usepackage{graphicx}
\usepackage{natbib}
\usepackage{scalerel}

\usepackage[table]{xcolor}

\usepackage{txfonts}
\usepackage[pdfencoding=auto,psdextra]{hyperref}
\hypersetup{
    colorlinks=true,
    linkcolor=blue,
    filecolor=magenta,      
    urlcolor=blue,
    citecolor=blue
}
\makeatletter
\renewcommand*\aa@pageof{, page \thepage{} of \pageref*{LastPage}}
\makeatother

\usepackage[utf8]{inputenc}

\usepackage[switch, modulo]{lineno}
\usepackage{amsmath}

\usepackage{amssymb}

\usepackage{subfigure}

\usepackage{chngcntr}
\usepackage{appendix}
\usepackage{etoolbox}
\usepackage{diagbox}
\usepackage{multicol}
\usepackage{comment}
\usepackage{siunitx}

\usepackage{multirow}
\usepackage{tabulary}
\usepackage[para]{threeparttable}
\usepackage{array,booktabs,longtable,tabularx}
\newcolumntype{L}{>{\raggedright\arraybackslash}X}
\usepackage{ltablex}

\definecolor{dodgerblue}{rgb}{0.12, 0.56, 1.0}
\definecolor{applegreen}{rgb}{0.55, 0.71, 0.0}
\definecolor{pink}{rgb}{0.858, 0.688, 0.688}

\usepackage{euclid}

\begin{document}
  \title{Euclid Quick Data Release (Q1)} \subtitle{Galaxy shapes and alignments in the cosmic web}


 
\newcommand{\orcid}[1]{} 
\author{Euclid Collaboration: C.~Laigle\orcid{0009-0008-5926-818X}\thanks{\email{laigle@iap.fr}}\inst{\ref{aff1}}
\and C.~Gouin\orcid{0000-0002-8837-9953}\inst{\ref{aff1}}
\and F.~Sarron\orcid{0000-0001-8376-0360}\inst{\ref{aff2},\ref{aff3}}
\and L.~Quilley\orcid{0009-0008-8375-8605}\inst{\ref{aff4}}
\and C.~Pichon\orcid{0000-0003-0695-6735}\inst{\ref{aff1},\ref{aff5}}
\and K.~Kraljic\orcid{0000-0001-6180-0245}\inst{\ref{aff6}}
\and F.~Durret\orcid{0000-0002-6991-4578}\inst{\ref{aff7}}
\and N.~.E.~Chisari\inst{\ref{aff8},\ref{aff9}}
\and U.~Kuchner\orcid{0000-0002-0035-5202}\inst{\ref{aff10}}
\and N.~Malavasi\orcid{0000-0001-9033-7958}\inst{\ref{aff11}}
\and M.~Magliocchetti\orcid{0000-0001-9158-4838}\inst{\ref{aff12}}
\and H.~J.~McCracken\orcid{0000-0002-9489-7765}\inst{\ref{aff1}}
\and J.~G.~Sorce\orcid{0000-0002-2307-2432}\inst{\ref{aff13},\ref{aff14}}
\and Y.~Kang\orcid{0009-0000-8588-7250}\inst{\ref{aff15}}
\and C.~J.~R.~McPartland\orcid{0000-0003-0639-025X}\inst{\ref{aff16},\ref{aff17}}
\and S.~Toft\orcid{0000-0003-3631-7176}\inst{\ref{aff18},\ref{aff17}}
\and N.~Aghanim\orcid{0000-0002-6688-8992}\inst{\ref{aff14}}
\and B.~Altieri\orcid{0000-0003-3936-0284}\inst{\ref{aff19}}
\and A.~Amara\inst{\ref{aff20}}
\and S.~Andreon\orcid{0000-0002-2041-8784}\inst{\ref{aff21}}
\and N.~Auricchio\orcid{0000-0003-4444-8651}\inst{\ref{aff22}}
\and H.~Aussel\orcid{0000-0002-1371-5705}\inst{\ref{aff23}}
\and C.~Baccigalupi\orcid{0000-0002-8211-1630}\inst{\ref{aff24},\ref{aff25},\ref{aff26},\ref{aff27}}
\and M.~Baldi\orcid{0000-0003-4145-1943}\inst{\ref{aff28},\ref{aff22},\ref{aff29}}
\and A.~Balestra\orcid{0000-0002-6967-261X}\inst{\ref{aff30}}
\and S.~Bardelli\orcid{0000-0002-8900-0298}\inst{\ref{aff22}}
\and A.~Basset\inst{\ref{aff31}}
\and P.~Battaglia\orcid{0000-0002-7337-5909}\inst{\ref{aff22}}
\and F.~Bernardeau\inst{\ref{aff32},\ref{aff1}}
\and A.~Biviano\orcid{0000-0002-0857-0732}\inst{\ref{aff25},\ref{aff24}}
\and A.~Bonchi\orcid{0000-0002-2667-5482}\inst{\ref{aff33}}
\and E.~Branchini\orcid{0000-0002-0808-6908}\inst{\ref{aff34},\ref{aff35},\ref{aff21}}
\and M.~Brescia\orcid{0000-0001-9506-5680}\inst{\ref{aff36},\ref{aff37}}
\and J.~Brinchmann\orcid{0000-0003-4359-8797}\inst{\ref{aff38},\ref{aff39}}
\and S.~Camera\orcid{0000-0003-3399-3574}\inst{\ref{aff40},\ref{aff41},\ref{aff42}}
\and G.~Ca\~nas-Herrera\orcid{0000-0003-2796-2149}\inst{\ref{aff43},\ref{aff44},\ref{aff9}}
\and V.~Capobianco\orcid{0000-0002-3309-7692}\inst{\ref{aff42}}
\and C.~Carbone\orcid{0000-0003-0125-3563}\inst{\ref{aff45}}
\and J.~Carretero\orcid{0000-0002-3130-0204}\inst{\ref{aff46},\ref{aff47}}
\and S.~Casas\orcid{0000-0002-4751-5138}\inst{\ref{aff48}}
\and M.~Castellano\orcid{0000-0001-9875-8263}\inst{\ref{aff49}}
\and G.~Castignani\orcid{0000-0001-6831-0687}\inst{\ref{aff22}}
\and S.~Cavuoti\orcid{0000-0002-3787-4196}\inst{\ref{aff37},\ref{aff50}}
\and K.~C.~Chambers\orcid{0000-0001-6965-7789}\inst{\ref{aff51}}
\and A.~Cimatti\inst{\ref{aff52}}
\and C.~Colodro-Conde\inst{\ref{aff53}}
\and G.~Congedo\orcid{0000-0003-2508-0046}\inst{\ref{aff54}}
\and C.~J.~Conselice\orcid{0000-0003-1949-7638}\inst{\ref{aff55}}
\and L.~Conversi\orcid{0000-0002-6710-8476}\inst{\ref{aff56},\ref{aff19}}
\and Y.~Copin\orcid{0000-0002-5317-7518}\inst{\ref{aff57}}
\and F.~Courbin\orcid{0000-0003-0758-6510}\inst{\ref{aff58},\ref{aff59}}
\and H.~M.~Courtois\orcid{0000-0003-0509-1776}\inst{\ref{aff60}}
\and M.~Cropper\orcid{0000-0003-4571-9468}\inst{\ref{aff61}}
\and A.~Da~Silva\orcid{0000-0002-6385-1609}\inst{\ref{aff62},\ref{aff63}}
\and H.~Degaudenzi\orcid{0000-0002-5887-6799}\inst{\ref{aff15}}
\and S.~de~la~Torre\inst{\ref{aff64}}
\and G.~De~Lucia\orcid{0000-0002-6220-9104}\inst{\ref{aff25}}
\and A.~M.~Di~Giorgio\orcid{0000-0002-4767-2360}\inst{\ref{aff12}}
\and C.~Dolding\orcid{0009-0003-7199-6108}\inst{\ref{aff61}}
\and H.~Dole\orcid{0000-0002-9767-3839}\inst{\ref{aff14}}
\and F.~Dubath\orcid{0000-0002-6533-2810}\inst{\ref{aff15}}
\and C.~A.~J.~Duncan\orcid{0009-0003-3573-0791}\inst{\ref{aff55}}
\and X.~Dupac\inst{\ref{aff19}}
\and A.~Ealet\orcid{0000-0003-3070-014X}\inst{\ref{aff57}}
\and S.~Escoffier\orcid{0000-0002-2847-7498}\inst{\ref{aff65}}
\and M.~Farina\orcid{0000-0002-3089-7846}\inst{\ref{aff12}}
\and R.~Farinelli\inst{\ref{aff22}}
\and F.~Faustini\orcid{0000-0001-6274-5145}\inst{\ref{aff33},\ref{aff49}}
\and S.~Ferriol\inst{\ref{aff57}}
\and F.~Finelli\orcid{0000-0002-6694-3269}\inst{\ref{aff22},\ref{aff66}}
\and S.~Fotopoulou\orcid{0000-0002-9686-254X}\inst{\ref{aff67}}
\and M.~Frailis\orcid{0000-0002-7400-2135}\inst{\ref{aff25}}
\and E.~Franceschi\orcid{0000-0002-0585-6591}\inst{\ref{aff22}}
\and S.~Galeotta\orcid{0000-0002-3748-5115}\inst{\ref{aff25}}
\and K.~George\orcid{0000-0002-1734-8455}\inst{\ref{aff68}}
\and W.~Gillard\orcid{0000-0003-4744-9748}\inst{\ref{aff65}}
\and B.~Gillis\orcid{0000-0002-4478-1270}\inst{\ref{aff54}}
\and C.~Giocoli\orcid{0000-0002-9590-7961}\inst{\ref{aff22},\ref{aff29}}
\and P.~G\'omez-Alvarez\orcid{0000-0002-8594-5358}\inst{\ref{aff69},\ref{aff19}}
\and J.~Gracia-Carpio\inst{\ref{aff11}}
\and B.~R.~Granett\orcid{0000-0003-2694-9284}\inst{\ref{aff21}}
\and A.~Grazian\orcid{0000-0002-5688-0663}\inst{\ref{aff30}}
\and F.~Grupp\inst{\ref{aff11},\ref{aff68}}
\and S.~Gwyn\orcid{0000-0001-8221-8406}\inst{\ref{aff70}}
\and S.~V.~H.~Haugan\orcid{0000-0001-9648-7260}\inst{\ref{aff71}}
\and H.~Hoekstra\orcid{0000-0002-0641-3231}\inst{\ref{aff9}}
\and W.~Holmes\inst{\ref{aff72}}
\and I.~M.~Hook\orcid{0000-0002-2960-978X}\inst{\ref{aff73}}
\and F.~Hormuth\inst{\ref{aff74}}
\and A.~Hornstrup\orcid{0000-0002-3363-0936}\inst{\ref{aff75},\ref{aff16}}
\and P.~Hudelot\inst{\ref{aff1}}
\and K.~Jahnke\orcid{0000-0003-3804-2137}\inst{\ref{aff76}}
\and M.~Jhabvala\inst{\ref{aff77}}
\and B.~Joachimi\orcid{0000-0001-7494-1303}\inst{\ref{aff78}}
\and E.~Keih\"anen\orcid{0000-0003-1804-7715}\inst{\ref{aff79}}
\and S.~Kermiche\orcid{0000-0002-0302-5735}\inst{\ref{aff65}}
\and A.~Kiessling\orcid{0000-0002-2590-1273}\inst{\ref{aff72}}
\and M.~Kilbinger\orcid{0000-0001-9513-7138}\inst{\ref{aff23}}
\and B.~Kubik\orcid{0009-0006-5823-4880}\inst{\ref{aff57}}
\and K.~Kuijken\orcid{0000-0002-3827-0175}\inst{\ref{aff9}}
\and M.~K\"ummel\orcid{0000-0003-2791-2117}\inst{\ref{aff68}}
\and M.~Kunz\orcid{0000-0002-3052-7394}\inst{\ref{aff80}}
\and H.~Kurki-Suonio\orcid{0000-0002-4618-3063}\inst{\ref{aff81},\ref{aff82}}
\and Q.~Le~Boulc'h\inst{\ref{aff83}}
\and A.~M.~C.~Le~Brun\orcid{0000-0002-0936-4594}\inst{\ref{aff84}}
\and D.~Le~Mignant\orcid{0000-0002-5339-5515}\inst{\ref{aff64}}
\and P.~Liebing\inst{\ref{aff61}}
\and S.~Ligori\orcid{0000-0003-4172-4606}\inst{\ref{aff42}}
\and P.~B.~Lilje\orcid{0000-0003-4324-7794}\inst{\ref{aff71}}
\and V.~Lindholm\orcid{0000-0003-2317-5471}\inst{\ref{aff81},\ref{aff82}}
\and I.~Lloro\orcid{0000-0001-5966-1434}\inst{\ref{aff85}}
\and G.~Mainetti\orcid{0000-0003-2384-2377}\inst{\ref{aff83}}
\and D.~Maino\inst{\ref{aff86},\ref{aff45},\ref{aff87}}
\and E.~Maiorano\orcid{0000-0003-2593-4355}\inst{\ref{aff22}}
\and O.~Mansutti\orcid{0000-0001-5758-4658}\inst{\ref{aff25}}
\and S.~Marcin\inst{\ref{aff88}}
\and O.~Marggraf\orcid{0000-0001-7242-3852}\inst{\ref{aff89}}
\and M.~Martinelli\orcid{0000-0002-6943-7732}\inst{\ref{aff49},\ref{aff90}}
\and N.~Martinet\orcid{0000-0003-2786-7790}\inst{\ref{aff64}}
\and F.~Marulli\orcid{0000-0002-8850-0303}\inst{\ref{aff91},\ref{aff22},\ref{aff29}}
\and R.~Massey\orcid{0000-0002-6085-3780}\inst{\ref{aff92}}
\and S.~Maurogordato\inst{\ref{aff93}}
\and E.~Medinaceli\orcid{0000-0002-4040-7783}\inst{\ref{aff22}}
\and S.~Mei\orcid{0000-0002-2849-559X}\inst{\ref{aff94},\ref{aff95}}
\and M.~Melchior\inst{\ref{aff96}}
\and Y.~Mellier\inst{\ref{aff7},\ref{aff1}}
\and M.~Meneghetti\orcid{0000-0003-1225-7084}\inst{\ref{aff22},\ref{aff29}}
\and E.~Merlin\orcid{0000-0001-6870-8900}\inst{\ref{aff49}}
\and G.~Meylan\inst{\ref{aff97}}
\and A.~Mora\orcid{0000-0002-1922-8529}\inst{\ref{aff98}}
\and M.~Moresco\orcid{0000-0002-7616-7136}\inst{\ref{aff91},\ref{aff22}}
\and L.~Moscardini\orcid{0000-0002-3473-6716}\inst{\ref{aff91},\ref{aff22},\ref{aff29}}
\and R.~Nakajima\orcid{0009-0009-1213-7040}\inst{\ref{aff89}}
\and C.~Neissner\orcid{0000-0001-8524-4968}\inst{\ref{aff99},\ref{aff47}}
\and S.-M.~Niemi\inst{\ref{aff43}}
\and J.~W.~Nightingale\orcid{0000-0002-8987-7401}\inst{\ref{aff100}}
\and C.~Padilla\orcid{0000-0001-7951-0166}\inst{\ref{aff99}}
\and S.~Paltani\orcid{0000-0002-8108-9179}\inst{\ref{aff15}}
\and F.~Pasian\orcid{0000-0002-4869-3227}\inst{\ref{aff25}}
\and K.~Pedersen\inst{\ref{aff101}}
\and W.~J.~Percival\orcid{0000-0002-0644-5727}\inst{\ref{aff102},\ref{aff103},\ref{aff104}}
\and V.~Pettorino\inst{\ref{aff43}}
\and S.~Pires\orcid{0000-0002-0249-2104}\inst{\ref{aff23}}
\and G.~Polenta\orcid{0000-0003-4067-9196}\inst{\ref{aff33}}
\and M.~Poncet\inst{\ref{aff31}}
\and L.~A.~Popa\inst{\ref{aff105}}
\and L.~Pozzetti\orcid{0000-0001-7085-0412}\inst{\ref{aff22}}
\and F.~Raison\orcid{0000-0002-7819-6918}\inst{\ref{aff11}}
\and R.~Rebolo\orcid{0000-0003-3767-7085}\inst{\ref{aff53},\ref{aff106},\ref{aff107}}
\and A.~Renzi\orcid{0000-0001-9856-1970}\inst{\ref{aff108},\ref{aff109}}
\and J.~Rhodes\orcid{0000-0002-4485-8549}\inst{\ref{aff72}}
\and G.~Riccio\inst{\ref{aff37}}
\and E.~Romelli\orcid{0000-0003-3069-9222}\inst{\ref{aff25}}
\and M.~Roncarelli\orcid{0000-0001-9587-7822}\inst{\ref{aff22}}
\and B.~Rusholme\orcid{0000-0001-7648-4142}\inst{\ref{aff110}}
\and R.~Saglia\orcid{0000-0003-0378-7032}\inst{\ref{aff68},\ref{aff11}}
\and Z.~Sakr\orcid{0000-0002-4823-3757}\inst{\ref{aff111},\ref{aff112},\ref{aff113}}
\and A.~G.~S\'anchez\orcid{0000-0003-1198-831X}\inst{\ref{aff11}}
\and D.~Sapone\orcid{0000-0001-7089-4503}\inst{\ref{aff114}}
\and B.~Sartoris\orcid{0000-0003-1337-5269}\inst{\ref{aff68},\ref{aff25}}
\and J.~A.~Schewtschenko\orcid{0000-0002-4913-6393}\inst{\ref{aff54}}
\and M.~Schirmer\orcid{0000-0003-2568-9994}\inst{\ref{aff76}}
\and P.~Schneider\orcid{0000-0001-8561-2679}\inst{\ref{aff89}}
\and T.~Schrabback\orcid{0000-0002-6987-7834}\inst{\ref{aff115}}
\and M.~Scodeggio\inst{\ref{aff45}}
\and A.~Secroun\orcid{0000-0003-0505-3710}\inst{\ref{aff65}}
\and G.~Seidel\orcid{0000-0003-2907-353X}\inst{\ref{aff76}}
\and M.~Seiffert\orcid{0000-0002-7536-9393}\inst{\ref{aff72}}
\and S.~Serrano\orcid{0000-0002-0211-2861}\inst{\ref{aff116},\ref{aff117},\ref{aff118}}
\and P.~Simon\inst{\ref{aff89}}
\and C.~Sirignano\orcid{0000-0002-0995-7146}\inst{\ref{aff108},\ref{aff109}}
\and G.~Sirri\orcid{0000-0003-2626-2853}\inst{\ref{aff29}}
\and J.~Skottfelt\orcid{0000-0003-1310-8283}\inst{\ref{aff119}}
\and L.~Stanco\orcid{0000-0002-9706-5104}\inst{\ref{aff109}}
\and J.~Steinwagner\orcid{0000-0001-7443-1047}\inst{\ref{aff11}}
\and P.~Tallada-Cresp\'{i}\orcid{0000-0002-1336-8328}\inst{\ref{aff46},\ref{aff47}}
\and A.~N.~Taylor\inst{\ref{aff54}}
\and H.~I.~Teplitz\orcid{0000-0002-7064-5424}\inst{\ref{aff120}}
\and I.~Tereno\inst{\ref{aff62},\ref{aff121}}
\and N.~Tessore\orcid{0000-0002-9696-7931}\inst{\ref{aff78}}
\and R.~Toledo-Moreo\orcid{0000-0002-2997-4859}\inst{\ref{aff122}}
\and F.~Torradeflot\orcid{0000-0003-1160-1517}\inst{\ref{aff47},\ref{aff46}}
\and I.~Tutusaus\orcid{0000-0002-3199-0399}\inst{\ref{aff112}}
\and L.~Valenziano\orcid{0000-0002-1170-0104}\inst{\ref{aff22},\ref{aff66}}
\and J.~Valiviita\orcid{0000-0001-6225-3693}\inst{\ref{aff81},\ref{aff82}}
\and T.~Vassallo\orcid{0000-0001-6512-6358}\inst{\ref{aff68},\ref{aff25}}
\and G.~Verdoes~Kleijn\orcid{0000-0001-5803-2580}\inst{\ref{aff123}}
\and A.~Veropalumbo\orcid{0000-0003-2387-1194}\inst{\ref{aff21},\ref{aff35},\ref{aff34}}
\and Y.~Wang\orcid{0000-0002-4749-2984}\inst{\ref{aff120}}
\and J.~Weller\orcid{0000-0002-8282-2010}\inst{\ref{aff68},\ref{aff11}}
\and A.~Zacchei\orcid{0000-0003-0396-1192}\inst{\ref{aff25},\ref{aff24}}
\and G.~Zamorani\orcid{0000-0002-2318-301X}\inst{\ref{aff22}}
\and F.~M.~Zerbi\inst{\ref{aff21}}
\and I.~A.~Zinchenko\orcid{0000-0002-2944-2449}\inst{\ref{aff68}}
\and E.~Zucca\orcid{0000-0002-5845-8132}\inst{\ref{aff22}}
\and V.~Allevato\orcid{0000-0001-7232-5152}\inst{\ref{aff37}}
\and M.~Ballardini\orcid{0000-0003-4481-3559}\inst{\ref{aff124},\ref{aff125},\ref{aff22}}
\and M.~Bolzonella\orcid{0000-0003-3278-4607}\inst{\ref{aff22}}
\and E.~Bozzo\orcid{0000-0002-8201-1525}\inst{\ref{aff15}}
\and C.~Burigana\orcid{0000-0002-3005-5796}\inst{\ref{aff126},\ref{aff66}}
\and R.~Cabanac\orcid{0000-0001-6679-2600}\inst{\ref{aff112}}
\and A.~Cappi\inst{\ref{aff22},\ref{aff93}}
\and D.~Di~Ferdinando\inst{\ref{aff29}}
\and J.~A.~Escartin~Vigo\inst{\ref{aff11}}
\and L.~Gabarra\orcid{0000-0002-8486-8856}\inst{\ref{aff127}}
\and M.~Huertas-Company\orcid{0000-0002-1416-8483}\inst{\ref{aff53},\ref{aff128},\ref{aff129},\ref{aff130}}
\and J.~Mart\'{i}n-Fleitas\orcid{0000-0002-8594-569X}\inst{\ref{aff98}}
\and S.~Matthew\orcid{0000-0001-8448-1697}\inst{\ref{aff54}}
\and N.~Mauri\orcid{0000-0001-8196-1548}\inst{\ref{aff52},\ref{aff29}}
\and R.~B.~Metcalf\orcid{0000-0003-3167-2574}\inst{\ref{aff91},\ref{aff22}}
\and A.~Pezzotta\orcid{0000-0003-0726-2268}\inst{\ref{aff131},\ref{aff11}}
\and M.~P\"ontinen\orcid{0000-0001-5442-2530}\inst{\ref{aff81}}
\and C.~Porciani\orcid{0000-0002-7797-2508}\inst{\ref{aff89}}
\and I.~Risso\orcid{0000-0003-2525-7761}\inst{\ref{aff132}}
\and V.~Scottez\inst{\ref{aff7},\ref{aff133}}
\and M.~Sereno\orcid{0000-0003-0302-0325}\inst{\ref{aff22},\ref{aff29}}
\and M.~Tenti\orcid{0000-0002-4254-5901}\inst{\ref{aff29}}
\and M.~Viel\orcid{0000-0002-2642-5707}\inst{\ref{aff24},\ref{aff25},\ref{aff27},\ref{aff26},\ref{aff134}}
\and M.~Wiesmann\orcid{0009-0000-8199-5860}\inst{\ref{aff71}}
\and Y.~Akrami\orcid{0000-0002-2407-7956}\inst{\ref{aff135},\ref{aff136}}
\and S.~Alvi\orcid{0000-0001-5779-8568}\inst{\ref{aff124}}
\and I.~T.~Andika\orcid{0000-0001-6102-9526}\inst{\ref{aff137},\ref{aff138}}
\and S.~Anselmi\orcid{0000-0002-3579-9583}\inst{\ref{aff109},\ref{aff108},\ref{aff139}}
\and M.~Archidiacono\orcid{0000-0003-4952-9012}\inst{\ref{aff86},\ref{aff87}}
\and F.~Atrio-Barandela\orcid{0000-0002-2130-2513}\inst{\ref{aff140}}
\and A.~Balaguera-Antolinez\orcid{0000-0001-5028-3035}\inst{\ref{aff53}}
\and C.~Benoist\inst{\ref{aff93}}
\and K.~Benson\inst{\ref{aff61}}
\and D.~Bertacca\orcid{0000-0002-2490-7139}\inst{\ref{aff108},\ref{aff30},\ref{aff109}}
\and M.~Bethermin\orcid{0000-0002-3915-2015}\inst{\ref{aff6}}
\and A.~Blanchard\orcid{0000-0001-8555-9003}\inst{\ref{aff112}}
\and L.~Blot\orcid{0000-0002-9622-7167}\inst{\ref{aff141},\ref{aff139}}
\and H.~B\"ohringer\orcid{0000-0001-8241-4204}\inst{\ref{aff11},\ref{aff142},\ref{aff143}}
\and S.~Borgani\orcid{0000-0001-6151-6439}\inst{\ref{aff144},\ref{aff24},\ref{aff25},\ref{aff26},\ref{aff134}}
\and M.~L.~Brown\orcid{0000-0002-0370-8077}\inst{\ref{aff55}}
\and S.~Bruton\orcid{0000-0002-6503-5218}\inst{\ref{aff145}}
\and A.~Calabro\orcid{0000-0003-2536-1614}\inst{\ref{aff49}}
\and B.~Camacho~Quevedo\orcid{0000-0002-8789-4232}\inst{\ref{aff116},\ref{aff118}}
\and F.~Caro\inst{\ref{aff49}}
\and C.~S.~Carvalho\inst{\ref{aff121}}
\and T.~Castro\orcid{0000-0002-6292-3228}\inst{\ref{aff25},\ref{aff26},\ref{aff24},\ref{aff134}}
\and F.~Cogato\orcid{0000-0003-4632-6113}\inst{\ref{aff91},\ref{aff22}}
\and T.~Contini\orcid{0000-0003-0275-938X}\inst{\ref{aff112}}
\and A.~R.~Cooray\orcid{0000-0002-3892-0190}\inst{\ref{aff146}}
\and O.~Cucciati\orcid{0000-0002-9336-7551}\inst{\ref{aff22}}
\and S.~Davini\orcid{0000-0003-3269-1718}\inst{\ref{aff35}}
\and F.~De~Paolis\orcid{0000-0001-6460-7563}\inst{\ref{aff147},\ref{aff148},\ref{aff149}}
\and G.~Desprez\orcid{0000-0001-8325-1742}\inst{\ref{aff123}}
\and A.~D\'iaz-S\'anchez\orcid{0000-0003-0748-4768}\inst{\ref{aff150}}
\and J.~J.~Diaz\inst{\ref{aff53}}
\and S.~Di~Domizio\orcid{0000-0003-2863-5895}\inst{\ref{aff34},\ref{aff35}}
\and J.~M.~Diego\orcid{0000-0001-9065-3926}\inst{\ref{aff151}}
\and P.-A.~Duc\orcid{0000-0003-3343-6284}\inst{\ref{aff6}}
\and A.~Enia\orcid{0000-0002-0200-2857}\inst{\ref{aff28},\ref{aff22}}
\and Y.~Fang\inst{\ref{aff68}}
\and A.~G.~Ferrari\orcid{0009-0005-5266-4110}\inst{\ref{aff29}}
\and P.~G.~Ferreira\orcid{0000-0002-3021-2851}\inst{\ref{aff127}}
\and A.~Finoguenov\orcid{0000-0002-4606-5403}\inst{\ref{aff81}}
\and A.~Fontana\orcid{0000-0003-3820-2823}\inst{\ref{aff49}}
\and A.~Franco\orcid{0000-0002-4761-366X}\inst{\ref{aff148},\ref{aff147},\ref{aff149}}
\and K.~Ganga\orcid{0000-0001-8159-8208}\inst{\ref{aff94}}
\and J.~Garc\'ia-Bellido\orcid{0000-0002-9370-8360}\inst{\ref{aff135}}
\and T.~Gasparetto\orcid{0000-0002-7913-4866}\inst{\ref{aff25}}
\and V.~Gautard\inst{\ref{aff152}}
\and E.~Gaztanaga\orcid{0000-0001-9632-0815}\inst{\ref{aff118},\ref{aff116},\ref{aff153}}
\and F.~Giacomini\orcid{0000-0002-3129-2814}\inst{\ref{aff29}}
\and F.~Gianotti\orcid{0000-0003-4666-119X}\inst{\ref{aff22}}
\and A.~H.~Gonzalez\orcid{0000-0002-0933-8601}\inst{\ref{aff154}}
\and G.~Gozaliasl\orcid{0000-0002-0236-919X}\inst{\ref{aff155},\ref{aff81}}
\and M.~Guidi\orcid{0000-0001-9408-1101}\inst{\ref{aff28},\ref{aff22}}
\and C.~M.~Gutierrez\orcid{0000-0001-7854-783X}\inst{\ref{aff156}}
\and A.~Hall\orcid{0000-0002-3139-8651}\inst{\ref{aff54}}
\and W.~G.~Hartley\inst{\ref{aff15}}
\and C.~Hern\'andez-Monteagudo\orcid{0000-0001-5471-9166}\inst{\ref{aff107},\ref{aff53}}
\and H.~Hildebrandt\orcid{0000-0002-9814-3338}\inst{\ref{aff157}}
\and J.~Hjorth\orcid{0000-0002-4571-2306}\inst{\ref{aff101}}
\and S.~Joudaki\orcid{0000-0001-8820-673X}\inst{\ref{aff46},\ref{aff153}}
\and J.~J.~E.~Kajava\orcid{0000-0002-3010-8333}\inst{\ref{aff158},\ref{aff159}}
\and V.~Kansal\orcid{0000-0002-4008-6078}\inst{\ref{aff160},\ref{aff161}}
\and D.~Karagiannis\orcid{0000-0002-4927-0816}\inst{\ref{aff124},\ref{aff162}}
\and K.~Kiiveri\inst{\ref{aff79}}
\and C.~C.~Kirkpatrick\inst{\ref{aff79}}
\and S.~Kruk\orcid{0000-0001-8010-8879}\inst{\ref{aff19}}
\and J.~Le~Graet\orcid{0000-0001-6523-7971}\inst{\ref{aff65}}
\and L.~Legrand\orcid{0000-0003-0610-5252}\inst{\ref{aff163},\ref{aff164}}
\and M.~Lembo\orcid{0000-0002-5271-5070}\inst{\ref{aff124},\ref{aff125}}
\and F.~Lepori\orcid{0009-0000-5061-7138}\inst{\ref{aff165}}
\and G.~Leroy\orcid{0009-0004-2523-4425}\inst{\ref{aff166},\ref{aff92}}
\and G.~F.~Lesci\orcid{0000-0002-4607-2830}\inst{\ref{aff91},\ref{aff22}}
\and J.~Lesgourgues\orcid{0000-0001-7627-353X}\inst{\ref{aff48}}
\and L.~Leuzzi\orcid{0009-0006-4479-7017}\inst{\ref{aff91},\ref{aff22}}
\and T.~I.~Liaudat\orcid{0000-0002-9104-314X}\inst{\ref{aff167}}
\and A.~Loureiro\orcid{0000-0002-4371-0876}\inst{\ref{aff168},\ref{aff169}}
\and J.~Macias-Perez\orcid{0000-0002-5385-2763}\inst{\ref{aff170}}
\and G.~Maggio\orcid{0000-0003-4020-4836}\inst{\ref{aff25}}
\and E.~A.~Magnier\orcid{0000-0002-7965-2815}\inst{\ref{aff51}}
\and F.~Mannucci\orcid{0000-0002-4803-2381}\inst{\ref{aff171}}
\and R.~Maoli\orcid{0000-0002-6065-3025}\inst{\ref{aff172},\ref{aff49}}
\and C.~J.~A.~P.~Martins\orcid{0000-0002-4886-9261}\inst{\ref{aff173},\ref{aff38}}
\and L.~Maurin\orcid{0000-0002-8406-0857}\inst{\ref{aff14}}
\and M.~Miluzio\inst{\ref{aff19},\ref{aff174}}
\and P.~Monaco\orcid{0000-0003-2083-7564}\inst{\ref{aff144},\ref{aff25},\ref{aff26},\ref{aff24}}
\and A.~Montoro\orcid{0000-0003-4730-8590}\inst{\ref{aff118},\ref{aff116}}
\and C.~Moretti\orcid{0000-0003-3314-8936}\inst{\ref{aff27},\ref{aff134},\ref{aff25},\ref{aff24},\ref{aff26}}
\and G.~Morgante\inst{\ref{aff22}}
\and S.~Nadathur\orcid{0000-0001-9070-3102}\inst{\ref{aff153}}
\and K.~Naidoo\orcid{0000-0002-9182-1802}\inst{\ref{aff153}}
\and A.~Navarro-Alsina\orcid{0000-0002-3173-2592}\inst{\ref{aff89}}
\and S.~Nesseris\orcid{0000-0002-0567-0324}\inst{\ref{aff135}}
\and F.~Passalacqua\orcid{0000-0002-8606-4093}\inst{\ref{aff108},\ref{aff109}}
\and K.~Paterson\orcid{0000-0001-8340-3486}\inst{\ref{aff76}}
\and L.~Patrizii\inst{\ref{aff29}}
\and A.~Pisani\orcid{0000-0002-6146-4437}\inst{\ref{aff65},\ref{aff175}}
\and D.~Potter\orcid{0000-0002-0757-5195}\inst{\ref{aff165}}
\and S.~Quai\orcid{0000-0002-0449-8163}\inst{\ref{aff91},\ref{aff22}}
\and M.~Radovich\orcid{0000-0002-3585-866X}\inst{\ref{aff30}}
\and P.-F.~Rocci\inst{\ref{aff14}}
\and S.~Sacquegna\orcid{0000-0002-8433-6630}\inst{\ref{aff147},\ref{aff148},\ref{aff149}}
\and M.~Sahl\'en\orcid{0000-0003-0973-4804}\inst{\ref{aff176}}
\and D.~B.~Sanders\orcid{0000-0002-1233-9998}\inst{\ref{aff51}}
\and E.~Sarpa\orcid{0000-0002-1256-655X}\inst{\ref{aff27},\ref{aff134},\ref{aff26}}
\and C.~Scarlata\orcid{0000-0002-9136-8876}\inst{\ref{aff177}}
\and A.~Schneider\orcid{0000-0001-7055-8104}\inst{\ref{aff165}}
\and M.~Schultheis\inst{\ref{aff93}}
\and D.~Sciotti\orcid{0009-0008-4519-2620}\inst{\ref{aff49},\ref{aff90}}
\and E.~Sellentin\inst{\ref{aff178},\ref{aff9}}
\and F.~Shankar\orcid{0000-0001-8973-5051}\inst{\ref{aff179}}
\and L.~C.~Smith\orcid{0000-0002-3259-2771}\inst{\ref{aff180}}
\and S.~A.~Stanford\orcid{0000-0003-0122-0841}\inst{\ref{aff181}}
\and K.~Tanidis\orcid{0000-0001-9843-5130}\inst{\ref{aff127}}
\and G.~Testera\inst{\ref{aff35}}
\and R.~Teyssier\orcid{0000-0001-7689-0933}\inst{\ref{aff175}}
\and S.~Tosi\orcid{0000-0002-7275-9193}\inst{\ref{aff34},\ref{aff35},\ref{aff21}}
\and A.~Troja\orcid{0000-0003-0239-4595}\inst{\ref{aff108},\ref{aff109}}
\and M.~Tucci\inst{\ref{aff15}}
\and C.~Valieri\inst{\ref{aff29}}
\and A.~Venhola\orcid{0000-0001-6071-4564}\inst{\ref{aff182}}
\and D.~Vergani\orcid{0000-0003-0898-2216}\inst{\ref{aff22}}
\and G.~Verza\orcid{0000-0002-1886-8348}\inst{\ref{aff183}}
\and P.~Vielzeuf\orcid{0000-0003-2035-9339}\inst{\ref{aff65}}
\and N.~A.~Walton\orcid{0000-0003-3983-8778}\inst{\ref{aff180}}
\and J.~R.~Weaver\orcid{0000-0003-1614-196X}\inst{\ref{aff184}}
\and D.~Scott\orcid{0000-0002-6878-9840}\inst{\ref{aff185}}}
										   
\institute{Institut d'Astrophysique de Paris, UMR 7095, CNRS, and Sorbonne Universit\'e, 98 bis boulevard Arago, 75014 Paris, France\label{aff1}
\and
Institut de Recherche en Informatique de Toulouse (IRIT), Universit\'e de Toulouse, CNRS, Toulouse INP, UT3, 31062 Toulouse, France\label{aff2}
\and
Laboratoire MCD, Centre de Biologie Int\'egrative (CBI), Universit\'e de Toulouse, CNRS, UT3, 31062 Toulouse, France\label{aff3}
\and
Centre de Recherche Astrophysique de Lyon, UMR5574, CNRS, Universit\'e Claude Bernard Lyon 1, ENS de Lyon, 69230, Saint-Genis-Laval, France\label{aff4}
\and
Kyung Hee University, Dept. of Astronomy \& Space Science, Yongin-shi, Gyeonggi-do 17104, Republic of Korea\label{aff5}
\and
Universit\'e de Strasbourg, CNRS, Observatoire astronomique de Strasbourg, UMR 7550, 67000 Strasbourg, France\label{aff6}
\and
Institut d'Astrophysique de Paris, 98bis Boulevard Arago, 75014, Paris, France\label{aff7}
\and
Institute for Theoretical Physics, Utrecht University, Princetonplein 5, 3584 CE Utrecht, The Netherlands\label{aff8}
\and
Leiden Observatory, Leiden University, Einsteinweg 55, 2333 CC Leiden, The Netherlands\label{aff9}
\and
School of Physics and Astronomy, University of Nottingham, University Park, Nottingham NG7 2RD, UK\label{aff10}
\and
Max Planck Institute for Extraterrestrial Physics, Giessenbachstr. 1, 85748 Garching, Germany\label{aff11}
\and
INAF-Istituto di Astrofisica e Planetologia Spaziali, via del Fosso del Cavaliere, 100, 00100 Roma, Italy\label{aff12}
\and
Univ. Lille, CNRS, Centrale Lille, UMR 9189 CRIStAL, 59000 Lille, France\label{aff13}
\and
Universit\'e Paris-Saclay, CNRS, Institut d'astrophysique spatiale, 91405, Orsay, France\label{aff14}
\and
Department of Astronomy, University of Geneva, ch. d'Ecogia 16, 1290 Versoix, Switzerland\label{aff15}
\and
Cosmic Dawn Center (DAWN), Denmark\label{aff16}
\and
Niels Bohr Institute, University of Copenhagen, Jagtvej 128, 2200 Copenhagen, Denmark\label{aff17}
\and
Cosmic Dawn Center (DAWN)\label{aff18}
\and
ESAC/ESA, Camino Bajo del Castillo, s/n., Urb. Villafranca del Castillo, 28692 Villanueva de la Ca\~nada, Madrid, Spain\label{aff19}
\and
School of Mathematics and Physics, University of Surrey, Guildford, Surrey, GU2 7XH, UK\label{aff20}
\and
INAF-Osservatorio Astronomico di Brera, Via Brera 28, 20122 Milano, Italy\label{aff21}
\and
INAF-Osservatorio di Astrofisica e Scienza dello Spazio di Bologna, Via Piero Gobetti 93/3, 40129 Bologna, Italy\label{aff22}
\and
Universit\'e Paris-Saclay, Universit\'e Paris Cit\'e, CEA, CNRS, AIM, 91191, Gif-sur-Yvette, France\label{aff23}
\and
IFPU, Institute for Fundamental Physics of the Universe, via Beirut 2, 34151 Trieste, Italy\label{aff24}
\and
INAF-Osservatorio Astronomico di Trieste, Via G. B. Tiepolo 11, 34143 Trieste, Italy\label{aff25}
\and
INFN, Sezione di Trieste, Via Valerio 2, 34127 Trieste TS, Italy\label{aff26}
\and
SISSA, International School for Advanced Studies, Via Bonomea 265, 34136 Trieste TS, Italy\label{aff27}
\and
Dipartimento di Fisica e Astronomia, Universit\`a di Bologna, Via Gobetti 93/2, 40129 Bologna, Italy\label{aff28}
\and
INFN-Sezione di Bologna, Viale Berti Pichat 6/2, 40127 Bologna, Italy\label{aff29}
\and
INAF-Osservatorio Astronomico di Padova, Via dell'Osservatorio 5, 35122 Padova, Italy\label{aff30}
\and
Centre National d'Etudes Spatiales -- Centre spatial de Toulouse, 18 avenue Edouard Belin, 31401 Toulouse Cedex 9, France\label{aff31}
\and
Institut de Physique Th\'eorique, CEA, CNRS, Universit\'e Paris-Saclay 91191 Gif-sur-Yvette Cedex, France\label{aff32}
\and
Space Science Data Center, Italian Space Agency, via del Politecnico snc, 00133 Roma, Italy\label{aff33}
\and
Dipartimento di Fisica, Universit\`a di Genova, Via Dodecaneso 33, 16146, Genova, Italy\label{aff34}
\and
INFN-Sezione di Genova, Via Dodecaneso 33, 16146, Genova, Italy\label{aff35}
\and
Department of Physics "E. Pancini", University Federico II, Via Cinthia 6, 80126, Napoli, Italy\label{aff36}
\and
INAF-Osservatorio Astronomico di Capodimonte, Via Moiariello 16, 80131 Napoli, Italy\label{aff37}
\and
Instituto de Astrof\'isica e Ci\^encias do Espa\c{c}o, Universidade do Porto, CAUP, Rua das Estrelas, PT4150-762 Porto, Portugal\label{aff38}
\and
Faculdade de Ci\^encias da Universidade do Porto, Rua do Campo de Alegre, 4150-007 Porto, Portugal\label{aff39}
\and
Dipartimento di Fisica, Universit\`a degli Studi di Torino, Via P. Giuria 1, 10125 Torino, Italy\label{aff40}
\and
INFN-Sezione di Torino, Via P. Giuria 1, 10125 Torino, Italy\label{aff41}
\and
INAF-Osservatorio Astrofisico di Torino, Via Osservatorio 20, 10025 Pino Torinese (TO), Italy\label{aff42}
\and
European Space Agency/ESTEC, Keplerlaan 1, 2201 AZ Noordwijk, The Netherlands\label{aff43}
\and
Institute Lorentz, Leiden University, Niels Bohrweg 2, 2333 CA Leiden, The Netherlands\label{aff44}
\and
INAF-IASF Milano, Via Alfonso Corti 12, 20133 Milano, Italy\label{aff45}
\and
Centro de Investigaciones Energ\'eticas, Medioambientales y Tecnol\'ogicas (CIEMAT), Avenida Complutense 40, 28040 Madrid, Spain\label{aff46}
\and
Port d'Informaci\'{o} Cient\'{i}fica, Campus UAB, C. Albareda s/n, 08193 Bellaterra (Barcelona), Spain\label{aff47}
\and
Institute for Theoretical Particle Physics and Cosmology (TTK), RWTH Aachen University, 52056 Aachen, Germany\label{aff48}
\and
INAF-Osservatorio Astronomico di Roma, Via Frascati 33, 00078 Monteporzio Catone, Italy\label{aff49}
\and
INFN section of Naples, Via Cinthia 6, 80126, Napoli, Italy\label{aff50}
\and
Institute for Astronomy, University of Hawaii, 2680 Woodlawn Drive, Honolulu, HI 96822, USA\label{aff51}
\and
Dipartimento di Fisica e Astronomia "Augusto Righi" - Alma Mater Studiorum Universit\`a di Bologna, Viale Berti Pichat 6/2, 40127 Bologna, Italy\label{aff52}
\and
Instituto de Astrof\'{\i}sica de Canarias, V\'{\i}a L\'actea, 38205 La Laguna, Tenerife, Spain\label{aff53}
\and
Institute for Astronomy, University of Edinburgh, Royal Observatory, Blackford Hill, Edinburgh EH9 3HJ, UK\label{aff54}
\and
Jodrell Bank Centre for Astrophysics, Department of Physics and Astronomy, University of Manchester, Oxford Road, Manchester M13 9PL, UK\label{aff55}
\and
European Space Agency/ESRIN, Largo Galileo Galilei 1, 00044 Frascati, Roma, Italy\label{aff56}
\and
Universit\'e Claude Bernard Lyon 1, CNRS/IN2P3, IP2I Lyon, UMR 5822, Villeurbanne, F-69100, France\label{aff57}
\and
Institut de Ci\`{e}ncies del Cosmos (ICCUB), Universitat de Barcelona (IEEC-UB), Mart\'{i} i Franqu\`{e}s 1, 08028 Barcelona, Spain\label{aff58}
\and
Instituci\'o Catalana de Recerca i Estudis Avan\c{c}ats (ICREA), Passeig de Llu\'{\i}s Companys 23, 08010 Barcelona, Spain\label{aff59}
\and
UCB Lyon 1, CNRS/IN2P3, IUF, IP2I Lyon, 4 rue Enrico Fermi, 69622 Villeurbanne, France\label{aff60}
\and
Mullard Space Science Laboratory, University College London, Holmbury St Mary, Dorking, Surrey RH5 6NT, UK\label{aff61}
\and
Departamento de F\'isica, Faculdade de Ci\^encias, Universidade de Lisboa, Edif\'icio C8, Campo Grande, PT1749-016 Lisboa, Portugal\label{aff62}
\and
Instituto de Astrof\'isica e Ci\^encias do Espa\c{c}o, Faculdade de Ci\^encias, Universidade de Lisboa, Campo Grande, 1749-016 Lisboa, Portugal\label{aff63}
\and
Aix-Marseille Universit\'e, CNRS, CNES, LAM, Marseille, France\label{aff64}
\and
Aix-Marseille Universit\'e, CNRS/IN2P3, CPPM, Marseille, France\label{aff65}
\and
INFN-Bologna, Via Irnerio 46, 40126 Bologna, Italy\label{aff66}
\and
School of Physics, HH Wills Physics Laboratory, University of Bristol, Tyndall Avenue, Bristol, BS8 1TL, UK\label{aff67}
\and
Universit\"ats-Sternwarte M\"unchen, Fakult\"at f\"ur Physik, Ludwig-Maximilians-Universit\"at M\"unchen, Scheinerstrasse 1, 81679 M\"unchen, Germany\label{aff68}
\and
FRACTAL S.L.N.E., calle Tulip\'an 2, Portal 13 1A, 28231, Las Rozas de Madrid, Spain\label{aff69}
\and
NRC Herzberg, 5071 West Saanich Rd, Victoria, BC V9E 2E7, Canada\label{aff70}
\and
Institute of Theoretical Astrophysics, University of Oslo, P.O. Box 1029 Blindern, 0315 Oslo, Norway\label{aff71}
\and
Jet Propulsion Laboratory, California Institute of Technology, 4800 Oak Grove Drive, Pasadena, CA, 91109, USA\label{aff72}
\and
Department of Physics, Lancaster University, Lancaster, LA1 4YB, UK\label{aff73}
\and
Felix Hormuth Engineering, Goethestr. 17, 69181 Leimen, Germany\label{aff74}
\and
Technical University of Denmark, Elektrovej 327, 2800 Kgs. Lyngby, Denmark\label{aff75}
\and
Max-Planck-Institut f\"ur Astronomie, K\"onigstuhl 17, 69117 Heidelberg, Germany\label{aff76}
\and
NASA Goddard Space Flight Center, Greenbelt, MD 20771, USA\label{aff77}
\and
Department of Physics and Astronomy, University College London, Gower Street, London WC1E 6BT, UK\label{aff78}
\and
Department of Physics and Helsinki Institute of Physics, Gustaf H\"allstr\"omin katu 2, 00014 University of Helsinki, Finland\label{aff79}
\and
Universit\'e de Gen\`eve, D\'epartement de Physique Th\'eorique and Centre for Astroparticle Physics, 24 quai Ernest-Ansermet, CH-1211 Gen\`eve 4, Switzerland\label{aff80}
\and
Department of Physics, P.O. Box 64, 00014 University of Helsinki, Finland\label{aff81}
\and
Helsinki Institute of Physics, Gustaf H{\"a}llstr{\"o}min katu 2, University of Helsinki, Helsinki, Finland\label{aff82}
\and
Centre de Calcul de l'IN2P3/CNRS, 21 avenue Pierre de Coubertin 69627 Villeurbanne Cedex, France\label{aff83}
\and
Laboratoire d'etude de l'Univers et des phenomenes eXtremes, Observatoire de Paris, Universit\'e PSL, Sorbonne Universit\'e, CNRS, 92190 Meudon, France\label{aff84}
\and
SKA Observatory, Jodrell Bank, Lower Withington, Macclesfield, Cheshire SK11 9FT, UK\label{aff85}
\and
Dipartimento di Fisica "Aldo Pontremoli", Universit\`a degli Studi di Milano, Via Celoria 16, 20133 Milano, Italy\label{aff86}
\and
INFN-Sezione di Milano, Via Celoria 16, 20133 Milano, Italy\label{aff87}
\and
University of Applied Sciences and Arts of Northwestern Switzerland, School of Computer Science, 5210 Windisch, Switzerland\label{aff88}
\and
Universit\"at Bonn, Argelander-Institut f\"ur Astronomie, Auf dem H\"ugel 71, 53121 Bonn, Germany\label{aff89}
\and
INFN-Sezione di Roma, Piazzale Aldo Moro, 2 - c/o Dipartimento di Fisica, Edificio G. Marconi, 00185 Roma, Italy\label{aff90}
\and
Dipartimento di Fisica e Astronomia "Augusto Righi" - Alma Mater Studiorum Universit\`a di Bologna, via Piero Gobetti 93/2, 40129 Bologna, Italy\label{aff91}
\and
Department of Physics, Institute for Computational Cosmology, Durham University, South Road, Durham, DH1 3LE, UK\label{aff92}
\and
Universit\'e C\^{o}te d'Azur, Observatoire de la C\^{o}te d'Azur, CNRS, Laboratoire Lagrange, Bd de l'Observatoire, CS 34229, 06304 Nice cedex 4, France\label{aff93}
\and
Universit\'e Paris Cit\'e, CNRS, Astroparticule et Cosmologie, 75013 Paris, France\label{aff94}
\and
CNRS-UCB International Research Laboratory, Centre Pierre Binetruy, IRL2007, CPB-IN2P3, Berkeley, USA\label{aff95}
\and
University of Applied Sciences and Arts of Northwestern Switzerland, School of Engineering, 5210 Windisch, Switzerland\label{aff96}
\and
Institute of Physics, Laboratory of Astrophysics, Ecole Polytechnique F\'ed\'erale de Lausanne (EPFL), Observatoire de Sauverny, 1290 Versoix, Switzerland\label{aff97}
\and
Aurora Technology for European Space Agency (ESA), Camino bajo del Castillo, s/n, Urbanizacion Villafranca del Castillo, Villanueva de la Ca\~nada, 28692 Madrid, Spain\label{aff98}
\and
Institut de F\'{i}sica d'Altes Energies (IFAE), The Barcelona Institute of Science and Technology, Campus UAB, 08193 Bellaterra (Barcelona), Spain\label{aff99}
\and
School of Mathematics, Statistics and Physics, Newcastle University, Herschel Building, Newcastle-upon-Tyne, NE1 7RU, UK\label{aff100}
\and
DARK, Niels Bohr Institute, University of Copenhagen, Jagtvej 155, 2200 Copenhagen, Denmark\label{aff101}
\and
Waterloo Centre for Astrophysics, University of Waterloo, Waterloo, Ontario N2L 3G1, Canada\label{aff102}
\and
Department of Physics and Astronomy, University of Waterloo, Waterloo, Ontario N2L 3G1, Canada\label{aff103}
\and
Perimeter Institute for Theoretical Physics, Waterloo, Ontario N2L 2Y5, Canada\label{aff104}
\and
Institute of Space Science, Str. Atomistilor, nr. 409 M\u{a}gurele, Ilfov, 077125, Romania\label{aff105}
\and
Consejo Superior de Investigaciones Cientificas, Calle Serrano 117, 28006 Madrid, Spain\label{aff106}
\and
Universidad de La Laguna, Departamento de Astrof\'{\i}sica, 38206 La Laguna, Tenerife, Spain\label{aff107}
\and
Dipartimento di Fisica e Astronomia "G. Galilei", Universit\`a di Padova, Via Marzolo 8, 35131 Padova, Italy\label{aff108}
\and
INFN-Padova, Via Marzolo 8, 35131 Padova, Italy\label{aff109}
\and
Caltech/IPAC, 1200 E. California Blvd., Pasadena, CA 91125, USA\label{aff110}
\and
Institut f\"ur Theoretische Physik, University of Heidelberg, Philosophenweg 16, 69120 Heidelberg, Germany\label{aff111}
\and
Institut de Recherche en Astrophysique et Plan\'etologie (IRAP), Universit\'e de Toulouse, CNRS, UPS, CNES, 14 Av. Edouard Belin, 31400 Toulouse, France\label{aff112}
\and
Universit\'e St Joseph; Faculty of Sciences, Beirut, Lebanon\label{aff113}
\and
Departamento de F\'isica, FCFM, Universidad de Chile, Blanco Encalada 2008, Santiago, Chile\label{aff114}
\and
Universit\"at Innsbruck, Institut f\"ur Astro- und Teilchenphysik, Technikerstr. 25/8, 6020 Innsbruck, Austria\label{aff115}
\and
Institut d'Estudis Espacials de Catalunya (IEEC),  Edifici RDIT, Campus UPC, 08860 Castelldefels, Barcelona, Spain\label{aff116}
\and
Satlantis, University Science Park, Sede Bld 48940, Leioa-Bilbao, Spain\label{aff117}
\and
Institute of Space Sciences (ICE, CSIC), Campus UAB, Carrer de Can Magrans, s/n, 08193 Barcelona, Spain\label{aff118}
\and
Centre for Electronic Imaging, Open University, Walton Hall, Milton Keynes, MK7~6AA, UK\label{aff119}
\and
Infrared Processing and Analysis Center, California Institute of Technology, Pasadena, CA 91125, USA\label{aff120}
\and
Instituto de Astrof\'isica e Ci\^encias do Espa\c{c}o, Faculdade de Ci\^encias, Universidade de Lisboa, Tapada da Ajuda, 1349-018 Lisboa, Portugal\label{aff121}
\and
Universidad Polit\'ecnica de Cartagena, Departamento de Electr\'onica y Tecnolog\'ia de Computadoras,  Plaza del Hospital 1, 30202 Cartagena, Spain\label{aff122}
\and
Kapteyn Astronomical Institute, University of Groningen, PO Box 800, 9700 AV Groningen, The Netherlands\label{aff123}
\and
Dipartimento di Fisica e Scienze della Terra, Universit\`a degli Studi di Ferrara, Via Giuseppe Saragat 1, 44122 Ferrara, Italy\label{aff124}
\and
Istituto Nazionale di Fisica Nucleare, Sezione di Ferrara, Via Giuseppe Saragat 1, 44122 Ferrara, Italy\label{aff125}
\and
INAF, Istituto di Radioastronomia, Via Piero Gobetti 101, 40129 Bologna, Italy\label{aff126}
\and
Department of Physics, Oxford University, Keble Road, Oxford OX1 3RH, UK\label{aff127}
\and
Instituto de Astrof\'isica de Canarias (IAC); Departamento de Astrof\'isica, Universidad de La Laguna (ULL), 38200, La Laguna, Tenerife, Spain\label{aff128}
\and
Universit\'e PSL, Observatoire de Paris, Sorbonne Universit\'e, CNRS, LERMA, 75014, Paris, France\label{aff129}
\and
Universit\'e Paris-Cit\'e, 5 Rue Thomas Mann, 75013, Paris, France\label{aff130}
\and
INAF - Osservatorio Astronomico di Brera, via Emilio Bianchi 46, 23807 Merate, Italy\label{aff131}
\and
INAF-Osservatorio Astronomico di Brera, Via Brera 28, 20122 Milano, Italy, and INFN-Sezione di Genova, Via Dodecaneso 33, 16146, Genova, Italy\label{aff132}
\and
ICL, Junia, Universit\'e Catholique de Lille, LITL, 59000 Lille, France\label{aff133}
\and
ICSC - Centro Nazionale di Ricerca in High Performance Computing, Big Data e Quantum Computing, Via Magnanelli 2, Bologna, Italy\label{aff134}
\and
Instituto de F\'isica Te\'orica UAM-CSIC, Campus de Cantoblanco, 28049 Madrid, Spain\label{aff135}
\and
CERCA/ISO, Department of Physics, Case Western Reserve University, 10900 Euclid Avenue, Cleveland, OH 44106, USA\label{aff136}
\and
Technical University of Munich, TUM School of Natural Sciences, Physics Department, James-Franck-Str.~1, 85748 Garching, Germany\label{aff137}
\and
Max-Planck-Institut f\"ur Astrophysik, Karl-Schwarzschild-Str.~1, 85748 Garching, Germany\label{aff138}
\and
Laboratoire Univers et Th\'eorie, Observatoire de Paris, Universit\'e PSL, Universit\'e Paris Cit\'e, CNRS, 92190 Meudon, France\label{aff139}
\and
Departamento de F{\'\i}sica Fundamental. Universidad de Salamanca. Plaza de la Merced s/n. 37008 Salamanca, Spain\label{aff140}
\and
Center for Data-Driven Discovery, Kavli IPMU (WPI), UTIAS, The University of Tokyo, Kashiwa, Chiba 277-8583, Japan\label{aff141}
\and
Ludwig-Maximilians-University, Schellingstrasse 4, 80799 Munich, Germany\label{aff142}
\and
Max-Planck-Institut f\"ur Physik, Boltzmannstr. 8, 85748 Garching, Germany\label{aff143}
\and
Dipartimento di Fisica - Sezione di Astronomia, Universit\`a di Trieste, Via Tiepolo 11, 34131 Trieste, Italy\label{aff144}
\and
California Institute of Technology, 1200 E California Blvd, Pasadena, CA 91125, USA\label{aff145}
\and
Department of Physics \& Astronomy, University of California Irvine, Irvine CA 92697, USA\label{aff146}
\and
Department of Mathematics and Physics E. De Giorgi, University of Salento, Via per Arnesano, CP-I93, 73100, Lecce, Italy\label{aff147}
\and
INFN, Sezione di Lecce, Via per Arnesano, CP-193, 73100, Lecce, Italy\label{aff148}
\and
INAF-Sezione di Lecce, c/o Dipartimento Matematica e Fisica, Via per Arnesano, 73100, Lecce, Italy\label{aff149}
\and
Departamento F\'isica Aplicada, Universidad Polit\'ecnica de Cartagena, Campus Muralla del Mar, 30202 Cartagena, Murcia, Spain\label{aff150}
\and
Instituto de F\'isica de Cantabria, Edificio Juan Jord\'a, Avenida de los Castros, 39005 Santander, Spain\label{aff151}
\and
CEA Saclay, DFR/IRFU, Service d'Astrophysique, Bat. 709, 91191 Gif-sur-Yvette, France\label{aff152}
\and
Institute of Cosmology and Gravitation, University of Portsmouth, Portsmouth PO1 3FX, UK\label{aff153}
\and
Department of Astronomy, University of Florida, Bryant Space Science Center, Gainesville, FL 32611, USA\label{aff154}
\and
Department of Computer Science, Aalto University, PO Box 15400, Espoo, FI-00 076, Finland\label{aff155}
\and
Instituto de Astrof\'\i sica de Canarias, c/ Via Lactea s/n, La Laguna 38200, Spain. Departamento de Astrof\'\i sica de la Universidad de La Laguna, Avda. Francisco Sanchez, La Laguna, 38200, Spain\label{aff156}
\and
Ruhr University Bochum, Faculty of Physics and Astronomy, Astronomical Institute (AIRUB), German Centre for Cosmological Lensing (GCCL), 44780 Bochum, Germany\label{aff157}
\and
Department of Physics and Astronomy, Vesilinnantie 5, 20014 University of Turku, Finland\label{aff158}
\and
Serco for European Space Agency (ESA), Camino bajo del Castillo, s/n, Urbanizacion Villafranca del Castillo, Villanueva de la Ca\~nada, 28692 Madrid, Spain\label{aff159}
\and
ARC Centre of Excellence for Dark Matter Particle Physics, Melbourne, Australia\label{aff160}
\and
Centre for Astrophysics \& Supercomputing, Swinburne University of Technology,  Hawthorn, Victoria 3122, Australia\label{aff161}
\and
Department of Physics and Astronomy, University of the Western Cape, Bellville, Cape Town, 7535, South Africa\label{aff162}
\and
DAMTP, Centre for Mathematical Sciences, Wilberforce Road, Cambridge CB3 0WA, UK\label{aff163}
\and
Kavli Institute for Cosmology Cambridge, Madingley Road, Cambridge, CB3 0HA, UK\label{aff164}
\and
Department of Astrophysics, University of Zurich, Winterthurerstrasse 190, 8057 Zurich, Switzerland\label{aff165}
\and
Department of Physics, Centre for Extragalactic Astronomy, Durham University, South Road, Durham, DH1 3LE, UK\label{aff166}
\and
IRFU, CEA, Universit\'e Paris-Saclay 91191 Gif-sur-Yvette Cedex, France\label{aff167}
\and
Oskar Klein Centre for Cosmoparticle Physics, Department of Physics, Stockholm University, Stockholm, SE-106 91, Sweden\label{aff168}
\and
Astrophysics Group, Blackett Laboratory, Imperial College London, London SW7 2AZ, UK\label{aff169}
\and
Univ. Grenoble Alpes, CNRS, Grenoble INP, LPSC-IN2P3, 53, Avenue des Martyrs, 38000, Grenoble, France\label{aff170}
\and
INAF-Osservatorio Astrofisico di Arcetri, Largo E. Fermi 5, 50125, Firenze, Italy\label{aff171}
\and
Dipartimento di Fisica, Sapienza Universit\`a di Roma, Piazzale Aldo Moro 2, 00185 Roma, Italy\label{aff172}
\and
Centro de Astrof\'{\i}sica da Universidade do Porto, Rua das Estrelas, 4150-762 Porto, Portugal\label{aff173}
\and
HE Space for European Space Agency (ESA), Camino bajo del Castillo, s/n, Urbanizacion Villafranca del Castillo, Villanueva de la Ca\~nada, 28692 Madrid, Spain\label{aff174}
\and
Department of Astrophysical Sciences, Peyton Hall, Princeton University, Princeton, NJ 08544, USA\label{aff175}
\and
Theoretical astrophysics, Department of Physics and Astronomy, Uppsala University, Box 515, 751 20 Uppsala, Sweden\label{aff176}
\and
Minnesota Institute for Astrophysics, University of Minnesota, 116 Church St SE, Minneapolis, MN 55455, USA\label{aff177}
\and
Mathematical Institute, University of Leiden, Einsteinweg 55, 2333 CA Leiden, The Netherlands\label{aff178}
\and
School of Physics \& Astronomy, University of Southampton, Highfield Campus, Southampton SO17 1BJ, UK\label{aff179}
\and
Institute of Astronomy, University of Cambridge, Madingley Road, Cambridge CB3 0HA, UK\label{aff180}
\and
Department of Physics and Astronomy, University of California, Davis, CA 95616, USA\label{aff181}
\and
Space physics and astronomy research unit, University of Oulu, Pentti Kaiteran katu 1, FI-90014 Oulu, Finland\label{aff182}
\and
Center for Computational Astrophysics, Flatiron Institute, 162 5th Avenue, 10010, New York, NY, USA\label{aff183}
\and
Department of Astronomy, University of Massachusetts, Amherst, MA 01003, USA\label{aff184}
\and
Department of Physics and Astronomy, University of British Columbia, Vancouver, BC V6T 1Z1, Canada\label{aff185}}

\date{}

\titlerunning{Galaxy shapes in the cosmic web with \Euclid}
   \authorrunning{Euclid Collaboration: C. Laigle et al.}

 \abstract{
 Galaxy morphologies and shape orientations are expected to correlate with their large-scale environment, since they grow by accreting matter from the cosmic web and are subject to interactions with other galaxies.
 Cosmic filaments are extracted in projection from the Euclid Quick Data Release~1 (covering 63.1~$\mathrm{deg}^2$) at  $0.5<z<0.9$  in tomographic slices of 170 comoving $h^{-1}\mathrm{Mpc}$ using photometric redshifts. 
 Galaxy morphologies are accurately retrieved thanks to the excellent resolution of VIS data.
 The distribution of massive galaxies ($M_* > 10^{10} M_\odot$) in the projected cosmic web is analysed as a function of morphology measured from VIS data. Specifically, the 2D alignment of galaxy shapes with large-scale filaments is quantified as a function of S\'ersic indices and masses.
 We find the known trend that more massive galaxies are closer to filament spines. At fixed stellar masses, morphologies 
 correlate both with densities and  distances to large-scale filaments.
 In addition, the large volume of this data set allows us to detect a signal indicating that there is a preferential alignment of the major axis of massive early-type galaxies along projected cosmic filaments. Overall, these results demonstrate our capabilities to carry out detailed studies of galaxy environments with \Euclid, which will be extended to higher redshift and lower stellar masses with the future Euclid Deep Survey. 
 }

\maketitle
\section{Introduction}
\label{sec:Intro}

Galaxies form from gas collapsing into the centres of dark matter (DM) halos \citep[see e.g.,][for a review]{Benson2010}, but these halos are not islands randomly distributed in the Universe.In a $\Lambda$CDM cosmology, it is now well established that the properties of  galaxies are determined to first order by the mass and  merger tree of their host DM halos, which in turn correlate with the amplitude of the local density \citep[the ``assembly bias", see e.g.,][]{1991ApJ...379..440B,Wechsler06, Alonso15}. The relation between star-formation rate and density (highlighted first in \citealp{1980ApJ...236..351D} and abundantly documented afterwards in the literature, see e.g., \citealp{malavasi17dens}, \citealp{Taamoli2024} for recent analyses) is a remarkable illustration of such a trend. 
On  larger scales the distribution of matter forms the  cosmic web \citep{Klypin1983,Bond1996}, a complex multi-scale network made of filaments and walls that border  regions of low density, called voids, and intersect in clusters of galaxies (see \citealp{Joeveer1978} for the report of the first detections and \citealp{deLapparent1986}, \citealp{geller89}  for  the  
identification of the
cosmic web in the large-scale distribution of galaxies). 
Such a web is theoretically and numerically expected to exist on many scales (from a few tens of kiloparsecs to several tens of megaparsecs). At large-scale the cosmic web emerges even from sparse tracers such as massive galaxies.

The large-scale cosmic filaments contain the material necessary to grow virialized structures: up to 40\% of the total matter \citep{cen06,aragoncalvo10} in the Universe and most of the gas at high redshift.
The thermodynamic properties of the gas in cosmic filaments and its kinematics is expected to regulate gas fuelling into galaxies, hence impacting their observable properties \citep[e.g.,][]{song21}. In addition, 
 the cosmic web anisotropy drives tides \citep{Codis2015,musso18,ramakrishnan19} and coherently shapes the angular momentum of large-scale flows \citep{pichon11,2013ApJ...769...74S,danovich15,laigle15}  impacting the accretion rate onto virialized structures,  the build-up of galaxy morphologies, and other secondary properties \citep[such that the dynamical state of the halos or their velocity anisotropy profiles, see e.g.,][]{2017MNRAS.469..594B}.

This role of the cosmic web on integrated galaxy  properties, such as mass, star-forming types, or morphology has now been assessed in observations at low redshift. A growing number of analyses \citep[e.g.,][]{Malavasi17,Kraljic18,Laigle2018} have provided evidence that the variations of these properties correlate not only with density, but also with distance to cosmic web filaments on various scales. In brief,  more massive galaxies are found closer to the filament spines. 
However, these results are still challenged, since limited statistics hamper their significance. It is still debated by some whether this trend can be entirely parametrised by a local density index \citep{Kane2024}, or if there is a second effect of proximity to the filament on top of the effect of local density \citep{Kuutma2017,Kraljic18}. In addition, it is so far observationally unclear how this signal evolves with redshift in terms of star-formation rate \citep[e.g.,][found an enhancement of star-forming galaxies in filaments at $z\simeq 1$]{darvish14} and morphology, since morphology was out of reach at high redshift from most past large-volume  photometric surveys. 
The first objective of this paper is to demonstrate with the Euclid Quick Data Release 1 \citep[Q1:][]{Q1-TP001} that the Euclid survey \citep{Laureijs2011,EuclidSkyOverview} will allow us to refine the analysis of the dependency of galaxy properties as a function of their large-scale environment. The Euclid survey will offer all together excellent galaxy morphologies, a very wide area, and a large redshift coverage. Although the Q1 data set is a very small fraction of the full Euclid Wide Survey (EWS), it covers about 30~times the area of the Cosmic Evolution Survey \citep[COSMOS,][]{COSMOS2007}, where studies on cosmic web were already successfully carried out \citep[see e.g.,][]{Darvish17,Laigle2018,Taamoli2024,Ko2024}.

It is theoretically expected that the imprint of the anisotropic cosmic web should be  stronger on galactic alignments than on masses and star-formation rates. Alignments are indeed related to the history of angular momentum acquisition, which is directly shaped by tides (the full tensor of the Hessian of the gravitational field), hence the \textit{anisotropic geometry} of the environment. The star-formation rate primarily depends on the host halo mass and growth \citep{Fu2025}, hence mostly on the \textit{amplitude} of the local density (the trace of the Hessian of the gravitational field).
Several theoretical studies have demonstrated that galaxy angular momenta and shapes are expected to be impacted by the proximity to their nearest larger-scale environment. These alignments are seeded in the initial conditions of the matter density field \citep{Codis2015}, and arise via tidal torquing \citep[][for a review]{Schafer2009}. Later on, matter accreted as secondary infall spins up  halos and rotation-dominated galaxies, which are expected to end up with their spins aligned with cosmic filaments up to a certain mass \citep[$\log_{10} M_{\rm *} /M_\odot < 10.5 $, see e.g.,][]{codis12,Codis2015,laigle15,Codis2018}.  Mergers along filaments  cause spin flip 
 \citep{codis12,Dubois2014,welker14,welker2016},
as orbital momentum is converted into internal momentum. 
This in turn suggests that bulge strength, a tracer for  merger remnant 
should correlate with alignment \citep{2022MNRAS.516.3569B}. 
As for massive dispersion-dominated objects, they are expected to be tidally stretched along the large-scale structure \citep{Schafer2009}. An alignment of galaxy spins and shapes is also expected with sheets at the boundary of voids \citep{dassignies22,davilakurban23}, but the observational signature of that is still unclear.  
The overall signal of alignment with cosmic web filaments has been widely explored in simulations \citep[e.g.,][]{Wang2018,GaneshaiahVeena2019,Lee2019,Lee2022,Kraljic2020,Cadiou2021,Cadiou2022b,Cadiou2022a}. 

Although these works agree qualitatively on the pattern of spin and shape alignments with filaments and sheets of the cosmic web, the amplitude of the trend and mass threshold for the spin transition are expected to depend on the scale of the filaments the galaxies live in \citep[e.g.,][]{codis12,Veena2021}. In addition,  gas accretion, mergers, and feedback are  highly non linear processes that will play a role in modifying the galaxy alignments signal. This is noticeable when comparing the alignment signal of galaxies measured in different simulations (e.g., comparing the {Horizon-AGN} simulation by \citealp{Dubois2014} and the Illustris-TNG simulation by \citealp{Nelson18}, \citealp{Pillepich18}), which assume various subgrid recipes \citep[the amplitude of the alignment signal with the density field can vary by a factor up to $2.4$ between different state-of-the-art hydrodynamical simulations, see][]{Zhang2023}. In this sense, if correctly understood, intrinsic alignments of specific populations of galaxies could directly be used as a constraint for galaxy-formation models. 

While intrinsic alignments offer valuable insights into galaxy formation and evolution, they also pose a significant challenge for weak gravitational lensing studies. Alignment between galaxy shapes and the cosmic web also results in pairwise alignments between neighbouring galaxies  (see e.g., \citealp{Joachimi2015} and \citealp{Lamman24} for reviews). This effect must be carefully quantified and modelled to avoid contamination on cosmological analyses based on cosmic shear statistics \citep{bernstein2002,hirata2004}. Based on simulations \cite{chisari15} and \cite{tenneti16} found that early-type galaxies tend to align radially (towards other early-type, or overdensities) while disc galaxies tend to align tangentially (see also, \citealp{samuroff21}; Euclid Collaboration: Hoffmann et al., in prep.). For disc galaxies, it is not clear if the alignment signal persists when measured in projection. 
Observationally, many detections of pairwise galaxy shape intrinsic alignments were reported for red and early-type galaxies \citep[see e.g.,][]{Singh2015, Georgiou19,Johnston2019,fortuna21,samuroff23,Hervas24} at low redshift, in a pattern qualitatively consistent with the simulation results, and dependent on wavelength and redshift. The detections are inevitably fewer at high redshift. \cite{TonegawaOkumura2022} reported the first detection of the  intrinsic alignment signal  at $z>1$ for red galaxies.
We emphasise that the present paper is not focused on measuring the pairwise intrinsic alignment signal for cosmological studies. We rather aim at demonstrating that quantifying the alignments of galaxy shapes with the cosmic web and their variations with galaxy masses and morphologies that will be possible with \Euclid, for the purpose of better constraining galaxy formation models.  However, the detection presented in this paper is promising for future mitigation of intrinsic alignment systematics in cosmic shear.

Several observational works already explored galaxy shape or spin alignments with respect either to cosmic filaments or large-scale overdensities. First alignment signals were reported in the Sloan Digital Sky Survey at $z<0.2$ \citep[SDSS, e.g,.][]{Jones2010,tempelstoica2013,tempellibeskind2013}, using the photometric shape of the galaxies as a proxy to infer the spin direction. Thanks to spectroscopy kinematics, spin alignment with filaments were also measured in the Mapping Nearby Galaxies at APO survey at $z<0.15$ \citep[MaNGA,][]{Kraljic2021}, in SAMI at $z<0.13$ \citep{Welker2020}, with the COSMOS HI Large Extragalactic Survey at $z<0.1$ \citep{BlueBird2020}. For galaxy shapes (of luminous red galaxies), alignments were measured in SDSS at $z\lesssim 0.4$ \citep[e.g.,][]{Singh2015,Chen2019,DesaiRyden2022}. Alignment of galaxies with the cosmic web filaments are in general inevitably restricted to low redshift, due to the difficulty to reconstruct the cosmic web environment at high redshift from photometric surveys. These studies consistently report shape alignments for red, elliptical galaxies (the galaxy major axes being aligned with cosmic filaments), but the alignment of blue, late-type galaxies is debated.  Recently, \cite{tonegawa24} reported a weak deviation from zero for the alignment of blue galaxies at $z\simeq 1.47$. We note that there is a specific interest of exploring galaxy alignments at higher redshift, because the late-type alignment signal is expected to strengthen at higher redshift \citep{Codis2018}, while the alignments of early-type should show the opposite trend. 

The Euclid survey will become a game changer in this respect, because it will offer both large volume and excellent imaging quality thanks to data from the VIS instrument \citep{EuclidSkyVIS} carefully processed by the OU-VIS pipeline \citep{Q1-TP002}. The second objective of this paper is therefore to demonstrate with the Q1 data set that the Euclid survey will allow us to place new constraints on these alignment trends.

Reconstructing the 3D cosmic web is anticipated to be feasible using the Euclid spectroscopic sample in the Euclid Deep Fields (EDFs) once they reach their full depth (Euclid Collaboration: Kraljic et al., in prep.). The Q1 data set, although covering the EDFs, is currently limited to the depth of the EWS. 
The density of tracers with reliable spectroscopic redshifts is consequently too sparse to provide an accurate view of the cosmic web at the scale required for galaxy evolution studies. Therefore, this paper explores an alternative method. Galaxy properties and orientations in the cosmic web will be studied in projection in thick tomographic slices using the photometric redshift sample, following a method sketched by \cite{Laigle2018} for the COSMOS field \citep[see also e.g.,][]{Sarron2019,Lazar23}. 

Throughout this paper, we adopt a Planck 2013 cosmology \citep{planck13}.
The paper is organised as follows. Section~\ref{sec:data} describes the observed and simulated data sets.  Section~\ref{sec:tool} presents
 the tool to extract the galaxy density field and the skeleton.  Section~\ref{sec:result} presents the measurements. Section~\ref{sec:conclusion}
wraps up and outlines future directions.

\section{Data}
\label{sec:data}

\subsection{Q1 release}
\label{sec:reldes}

This work relies on the photometric catalogues from the Q1 data set, which have been extracted from the three EDFs: EDF North (EDF-N), EDF South (EDF-S) and EDF Fornax (EDF-F). These are described in \cite{Q1-TP001}. Although they span the EDFs, the Q1 data set are representative of the EWS in terms of photometric depth, namely a $5\,\sigma$ depth of 26.0, 23.8, 24.0, 24.0 in the $\IE$, $\YE$, $\JE$, and $\HE$ filters, respectively \citep{Q1-TP005}. Photometric catalogues have been extracted by the OU-MER pipeline \citep{Q1-TP004} from images from the VIS \citep[$\IE$,][]{EuclidSkyVIS} and NISP \citep[$\YE$, $\JE$, $\HE$,][]{EuclidSkyNISP} instruments processed by OU-VIS \citep{Q1-TP002} and OU-NIR \citep{Q1-TP003}, together with external companion data from ground-based telescopes. Although the area of the Q1 data is very small with respect to existing stage-III surveys, such as the Dark Energy Survey \citep[][]{DES2015}, the Kilo-Degree Survey \citep[][]{Kids2015}, and the Hyper Suprime-Cam Subaru Strategic Program \citep[][]{Aihara2018}, it is competitive in terms of depth, and the VIS images have much better resolution (see below), which is crucial to measure accurate morphologies and shape orientations.

\subsubsection{Galaxy morphologies}
VIS images are particularly valuable to derive accurate galaxy morphologies and orientations on the sky, thanks to the pixel size of $\ang{;;0.1}$ 
and excellent point spread function \citep[PSF, with a full width at half maximum smaller than $\ang{;;0.16}$, see][]{Q1-TP002}. Catalogues from OU-MER contain a variety of morphological measurements for detected parameters, including S\'ersic parametric fits \citep{sersic63} measured with the \texttt{SourceXtractor++} software \citep{Bertin-2020-SourceXtractor-plus-plus, Kummel-2022-SE-use}, and deep-learning-based visual-like morphologies \citep{Q1-TP004}. In this work, we rely in particular on the S\'ersic index measured on the VIS images  through \texttt{SourceXtractor++} to quantify galaxy morphologies \citep[see][for a comprehensive study of galaxy morphologies in Q1 data]{Q1-SP040}. We consider massive galaxies at low redshift, which are mostly very well resolved (see Fig.~\ref{fig:propsample}), and therefore have very robust S\'ersic indices (with a typical error in $n_{\rm sersic}$ smaller than 0.2).

\subsubsection{Photometric redshifts and stellar masses}
\label{subsec:photoz}
Photometric redshifts and stellar masses are obtained through the OU-PHZ pipeline \citep{Q1-TP005}. OU-PHZ has developed two methods for photometric redshift estimates, one based on template-fitting called {\tt Phosphoros}, and another one based on machine-learning called {Nearest-Neighbour Photometric Redshifts ({\tt NNPZ})}. As described in \cite{Q1-TP005}, in the Q1 release, {\tt Phosphoros} is used to derive photometric redshifts and return the Bayesian posterior distribution of the redshift for each galaxy. On the other hand, {\tt NNPZ} is a supervised-learning algorithm that computes galaxy properties (redshift, masses, and other properties) based on the nearest-neighbours method. It therefore requires a calibration sample. The Q1 release of  {\tt NNPZ} properties contains the mode, median and 68\% percentiles of the 30 closest weighted neighbours in the reference sample, but no Bayesian posteriors. We note that galaxies are therefore given a redshift both from the {\tt Phosphoros} and {\tt NNPZ} algorithms. In this paper, we need to rely on the redshift posteriors to build many realisations of our tomographic slices, as described in Sect.~\ref{sec:slices}. We therefore rely on the {\tt Phosphoros} output {to reconstruct the cosmic web}, since the full posterior from {\tt NNPZ} is not available. 
For consistency, we use also the {\tt Phosphoros} photometric redshifts to {assign galaxies} to their most likely host slices. However, to quantify the evolution of galaxy masses in the cosmic web, we use the  {\tt NNPZ} stellar masses, given that masses are not available from the {\tt Phosphoros} algorithm. As a consequence, we keep in our final fiducial sample only galaxies for which the  median {\tt NNPZ} redshift is included in the smallest interval containing 70\% (i.e., encompassing $\pm 1\sigma$, this interval is an output of the Q1 release) of the {\tt Phosphoros} redshift posterior distribution of a galaxy, called PDF($z$). The percentage of galaxies for which the median {\tt NNPZ} redshift does not fall in the 70\% interval from the {\tt Phosphoros} PDF($z$) is higher in EDF-N, with values of 30.7\%, 24.3\%, and 23.9\% 
in EDF-N, EDF-S and EDF-F, respectively 
(all masses, assuming that the mode of the {\tt Phosphoros} redshift and the median {\tt NNPZ} redshift are both in the range 0.5--0.9). By comparing to spectroscopic redshifts available through public surveys, \cite{Q1-TP005} found  a dispersion $\sigma_{\rm NMAD}\simeq 0.03$ and a fraction of catastrophic outliers $\eta\simeq 0.1$ at $\IE<23$ in all EDFs. At $\IE<24.5$, the  dispersion is $\sigma_{\rm NMAD}\simeq 0.04$ and $\sigma_{\rm NMAD}\simeq 0.06$ and the fraction of catastrophic outliers is $\eta\simeq 0.16$  and $\eta\simeq 0.27$  in the EDF-N and EDF-F+EDF-S respectively.

\subsubsection{Final sample}
\label{sec:finsamp}
The catalogues used in this study were retrieved through the Euclid science archive system, 
selecting all galaxies from the MER catalogue with the mode of the photometric redshift from {\tt Phosphoros} $z<1$. We then apply a series of cuts to our final sample to maximise its quality and remove potential artefacts \citep[see also][for a full explanation on these cuts]{Q1-SP040}. Figure~\ref{fig:propsample} presents the properties of this sample in comparison to the VIS depth and PSF FWHM:
\begin{itemize}
    \item {\tt VIS\_DET}$=1$ (the source must be detected in the $\IE$ band) and {\tt flux\_vis\_Sersic }$> 0$ (the source must have a non-null flux from the Sersic fit);
    \item {\tt phz\_flags }$=0$ and {\tt phys\_param\_flags}$=0$ (the source has been classified as a galaxy by the OU-PHZ pipeline);
\item ${\mathrm{\tt Sersic\_Sersic\_vis\_axis\_ratio} > 0.05}$ together with ${\mathrm{\tt Sersic\_Sersic\_vis\_radius} < 2\, \mathrm{\tt semimajor\_axis}}$, which mostly remove residual diffraction spikes of bright stars or cosmic rays;
\item  {\tt Sersic\_Sersic\_vis\_radius} $>\ang{;;0.16}$, where $\ang{;;0.16}$ is taken as the VIS PSF FWHM \citep{Q1-TP002}, to remove unresolved galaxies. 
\end{itemize}

When analysing galaxy properties, we impose additional conditions on  their redshift, as explained in Sect.~\ref{subsec:photoz}. We also impose conditions on the S\'ersic parameters to avoid fits that reach the limits of the parameter space instead of converging to a meaningful solution:
{\tt Sersic\_Sersic\_vis\_axis\_ratio }$<1.0$ and 
$0.302 <$ {\tt Sersic\_Sersic\_vis\_index} $< 5.45$.
When measuring galaxy shape orientations with respect to filaments, we  select galaxies with a certain elongation, required to measure an angle with respect to filaments, by imposing {\tt Sersic\_Sersic\_vis\_axis\_ratio} $<$ 0.9. We also impose the error on the position angle {\tt Sersic\_angle\_err} to remain below~1.  Because we rely on the S\'ersic index to separate early-type and late-type galaxies, we also impose the error on the S\'ersic index {\tt Sersic\_err} to be smaller than 1. We discuss in Fig.~\ref{fig:prop_selection} to what extent these cuts bias our fiducial sample. Although these cuts guarantee a clean sample, we note that relaxing them does not change our conclusions (see Fig.~\ref{fig:variations_selection}).

\begin{figure}
    \centering    \includegraphics[width=0.9\columnwidth]{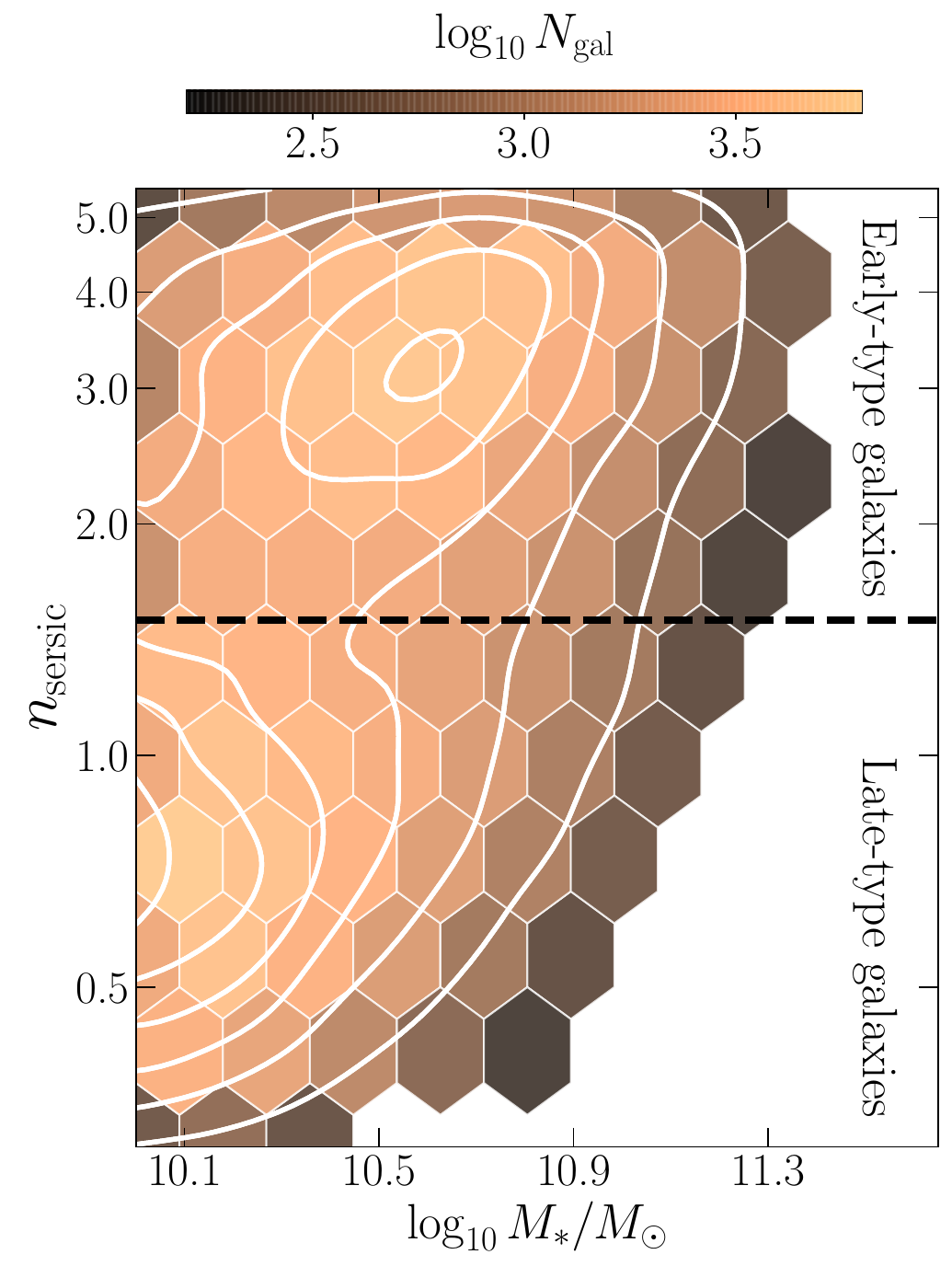}
    \caption{Galaxy distribution in the plane of S\'ersic index versus mass for all galaxies in our fiducial sample in the EDFs at $0.5<z<0.9$. The black dashed line corresponds to the boundary for the early-type galaxy domain. Density contours are overlaid in white.}
    \label{fig:bimodality}
\end{figure}

\subsubsection{Early-type galaxy population}
\label{sec:ET}
Figure~\ref{fig:bimodality} displays the galaxy distribution in the plane of S\'ersic index versus stellar mass. Only hexagons containing more than 200 galaxies are displayed. The S\'ersic index characterises the steepness of the light profile. Early-type galaxies tend to have a more concentrated light profile, hence a higher S\'ersic index. Therefore this representation allows us to easily distinguish between early-type and late-type galaxies \citep{Q1-SP040}. Early-type galaxies are those located at the top of the diagram, with mostly $\log_{10} (M_*/M_\odot)>10.3$ and S\'ersic index $n_{\rm sersic}>1.5$. We therefore adopt in our analysis the criterion of $n_{\rm sersic}>1.5$ for this population, with this S\'ersic index threshold roughly corresponding to a minimum in the isocontours on this diagram (Fig.~\ref{fig:alignment_sersicdiff} highlights how our alignment result changes when varying this criterion).

\begin{figure*}
\centering\includegraphics[width=0.95\textwidth]{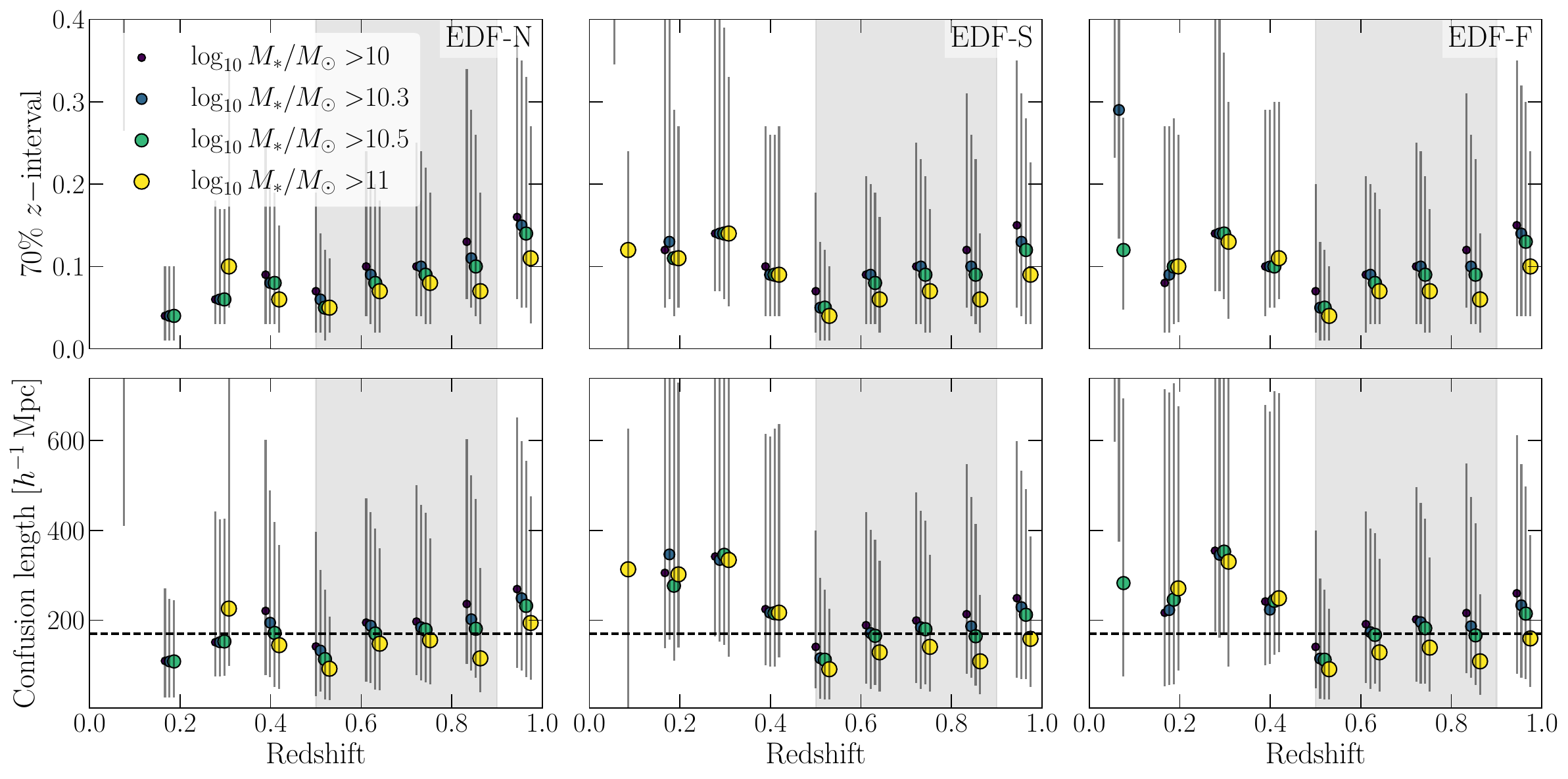}
\caption{\emph{Top panels}: smallest redshift interval containing $70$\% of the probability derived from the {\tt Phosphoros} PDF($z$) as a function of redshift and masses for galaxies in EDF-N (left), EDF-S (middle), and EDF-F (right), as described in Sect.~\ref{sec:reldes}. In each redshift bin, the markers indicate the mean of the $70$\% $z$-interval, and the error bars indicate the 16th and 84th percentiles. A $5\,\sigma$ clipping is applied before computing these statistics. Because we rely on the {\tt NNPZ} stellar masses to select galaxies, we also impose for consistency that the  {\tt NNPZ} median redshift fall in the $70$\% $z$-interval.  \emph{Bottom panels}: the corresponding uncertainty in $h^{-1}{\rm Mpc}$, which characterises the typical confusion length of galaxies along the line of sight. The black dashed line indicates the thickness for tomographic slices that we consider in this study, calibrated to encompass photometric redshift errors of galaxies more massive than $10^{10.3}M_\odot$ at $0.5<z<0.9$. The grey shaded area indicates the redshift range considered in this study.
}
\label{fig:confusion_length}
\end{figure*}

\begin{figure*}
\centering\includegraphics[width=0.99\textwidth]{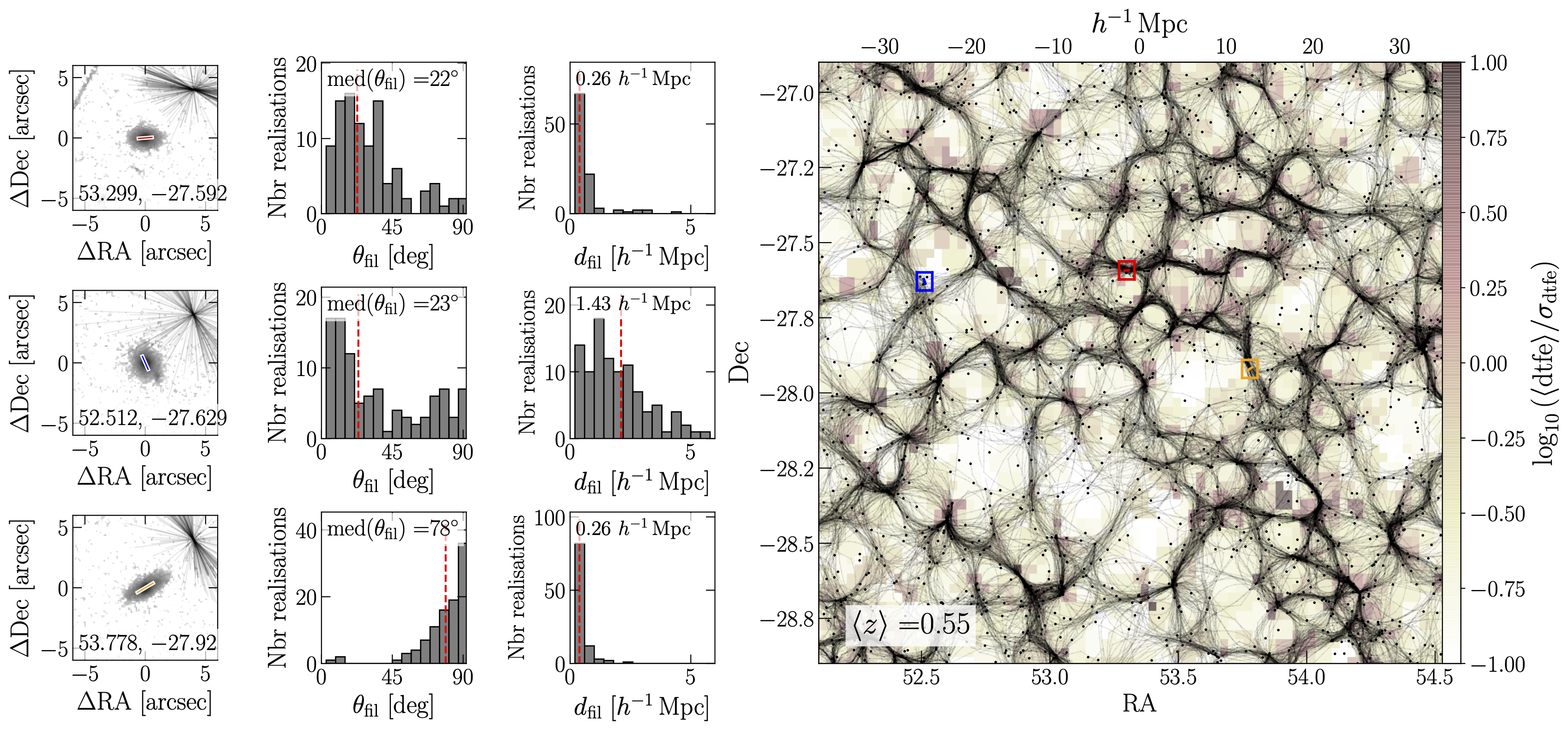}
\caption{Zoom into the slice from the EDF-F at $z=0.55$ with three selected galaxy cutouts from the MER mosaic (left subpanels), with their positions in the slice marked with coloured squares. In the top right corner of each cutout, we displayed the orientation of the closest segment in each of the realisations (the distance to the galaxy on the stamp is not representative of the real distance). The coordinates of the galaxies are indicated in the bottom of the cutout. The slice thickness is $170$~$h^{-1}{\rm Mpc}$. One hundred realisations from the \texttt{DisPerSE} skeleton extracted in 2D are overplotted, color-coded by $\log_{10}\left[\langle {\rm dtfe}\rangle/\sigma_{\rm dtfe}\right]$, which is the density estimation from the Delaunay tesselation (in one of the realisations) in the background (normalised by the standard deviation). Dots represent galaxies in one realisation of the slice, with their size scaling with galaxy masses.  The subpanels in the two middle columns display the distribution of galaxy projected angles and distances between the galaxy main axis and the closest filament in each of the 100 realisations. We made use of the ESA Datalabs resources \citep{Datalabscite} to extract galaxy stamps shown here.
}
\label{fig:method}
\end{figure*}

\subsection{The DAWN catalogues}
For the purpose of understanding the impact of photometric redshift uncertainties on the reconstructed cosmic filaments, we rely on the DR1  catalogues \citep{EP-Zalesky} from the Cosmic Dawn Survey \citep[DAWN,][]{EP-McPartland}, which are made 
available for part of two EDFs: Euclid Deep Field North and Euclid Deep Field
Fornax. These catalogues (photometry extraction and computation of galaxy properties) were derived independently of \Euclid: therefore we do not use them to derive the main results presented in this work. However, in particular thanks to the deep optical imaging from Hyper Suprime Camera on Subaru \citep{Miyazaki2018} and  the IR photometry from \textit{Spitzer}/IRAC data \citep{Moneti-EP17}, this data set reaches higher photometric redshift accuracy than the Q1 catalogues, especially concerning the fraction of catastrophic outliers (with a dispersion $\sigma_{\rm NMAD}\simeq 0.05$ and a fraction of catastrophic outliers $\eta<0.08$ at $i<25$, for photometric redshifts derived with the \texttt{LePhare} software, \citealp{ilbert2006,LePHARE2011}). They can be used to validate our cosmic-web extraction. This quality assessment is provided in Appendix~\ref{sec:qualassess}. In addition, Fig.~\ref{fig:variations_skeleton} shows the alignment results derived using these catalogues. 

\section{Methods}
\label{sec:tool}

\subsection{Designing tomographic slices}

\label{sec:slices}

We rely on photometric redshifts derived with {\tt Phosphoros} to estimate the position of galaxies along the line of sight. The smallest redshift interval encompassing 70$\%$ of the PDF($z$) of galaxies more massive than $10^{10}  M_\odot$ is of the order of 0.1 at $z<1$ (see Fig.~\ref{fig:confusion_length}), which makes a 3D-reconstruction of the cosmic web network impossible. 
Therefore, cosmic filaments are extracted in 2D tomographic slices, the thickness of which is calibrated on the typical redshift uncertainties \citep[see e.g.,][]{Laigle2018}. The {top} panels of Fig.~\ref{fig:confusion_length} presents the redshift uncertainties up to $z= 1$ in the three EDFs for different mass thresholds. The  smallest redshift interval encompassing 70$\%$ of the PDF($z$) were derived from the {\tt Phosphoros} output, as a function of redshift and galaxy masses. In each redshift bin, the markers indicate the mean of the redshift interval, and the error bars indicate the 16th and 84th percentiles of the galaxy distribution in this bin. A $5\,\sigma$ clipping is applied before computing these statistics. 
The bottom panel translates this uncertainty in terms of  the equivalent confusion length in comoving $h^{-1}{\rm Mpc}$. At $0.5<z<0.9$, the corresponding comoving length  is of the order of 170~$h^{-1}{\rm Mpc}$ for galaxies more massive than $10^{10.3}M_\odot$. We exclude the range $z<0.5$ where the lack of blue optical photometry does not allow us to precisely constrain precisely the Balmer break for these galaxies, hence limiting the redshift accuracy \citep[see e.g.,][]{laigle2019}. We therefore extract the cosmic web in slices of thickness 170~$h^{-1}{\rm Mpc}$  using only galaxies more massive than $10^{10.3}M_\odot$, with an overlap thickness of $90$~$h^{-1}{\rm Mpc}$. We have 10 overlapping slices over $0.5<z<0.9$.

Given their photometric redshift uncertainties, galaxies only have a certain probability to be present in a given tomographic slice. To account for this uncertainty in the reconstruction, we perform 100~realisations of each slice by drawing redshifts under the PDF($z$) of each galaxy. We note that the density of galaxies used as tracers of the cosmic web  varies slightly over the redshift range, owing both to the  evolution of the galaxy number density above this mass threshold, and to the widening of the galaxy redshift posterior distribution towards higher redshift.
We choose not to correct for this evolving density, since disentangling the underlying causes is not straightforward. Given the minimal evolution of the massive end of the mass function over this redshift range \citep[see e.g.,][]{Shuntov2022}, this choice is reasonable. However, this approach will need to be revisited when extending the redshift range with future Euclid releases.

\subsection{Filament identifications}
\label{sec:cwextraction}

The cosmic web is extracted using the \texttt{DisPerSE} algorithm \citep{sousbie11a,sousbie11b}. The density field is first estimated in each realisation of each slice by computing a Delaunay tessellation \citep{Delaunay_1934aa} on the discrete distribution of galaxies. The angular sky coordinates of galaxies were converted to their stereographic projection prior to this computation.\footnote{The stereographic projection does not change the triangulation (the way galaxies are connected either on the sphere surface or in the projection is the same), but does change the area of triangles. In \texttt{DisPerSE}, the density is taken as inversely proportional to the area of a triangle in the tessellation.  Given that the extents of the fields are small (a few degrees at most), the ratio between the area of the triangles computed on the sphere (in 3D cartesian coordinates) and computed on the projection varies very little across the field. We checked that whether or not we replace the area of the triangle in the tessellation by the one it would have on the sphere has no impact on the resulting filament catalogue.}
Cosmic filaments are then identified in 2D in each realisation of each slice as the special lines connecting topologically robust pairs of saddle-peak critical points. Therefore, the extraction is non-local in the sense that what determines the presence of a filament at a given location is not only the amplitude of the overdensity with respect to the local background, but also the distribution of matter on a larger scale. 
The set of segments that compose the extracted cosmic web is called the ``skeleton" in the following.

\subsection{Persistence} 
In \texttt{DisPerSE}, the extraction of cosmic structures is parameterised by the persistence threshold. Persistence measures the significance of critical pairs within the Delaunay tessellation built on a random discrete Poisson distribution, and is expressed in numbers of $\sigma$. Thus, the filtering of low-persistence structures
ensures that the extraction is robust with respect to noise.
Previous studies calibrated the persistence threshold relying on comparisons with mocks extracted from cosmological simulations. In the context of \Euclid,  Euclid Collaboration: Malavasi et al., in prep. investigated with mocks mimicking the noise level of the EWS that such a low persistence threshold (of the order of $2\,\sigma$) is required in 2D to guarantee the retrieval of cosmic structures that are at a persistence of the order of $5\,\sigma$ in 3D (i.e., that are very significant with respect to noise).
The main results of this work are based on a persistent threshold of $1.5\,\sigma$ to maximise the number of detected structures. We show in Appendix~\ref{app:qualassessangle} that choosing a $2\,\sigma$ persistence threshold does not change our conclusions. Notably our approach inherently filters out unreliable filaments by performing multiple realizations of the cosmic web extraction through sampling the PDF($z$) of galaxies. Robust filaments are those that consistently emerge from the noise across many realisations, as illustrated in Fig.~\ref{fig:method}.

\subsection{Smoothing}
Filaments extracted by \texttt{DisPerSE} are composed of segments that follow the edges of triangles in the Delaunay tessellation. Their direction is therefore stochastic at very small scales due to the discretisation of the particle distribution that traces the density field. To capture the main direction of filaments and accurately measure the alignment of galaxy shapes, the skeleton is smoothed over 10 segments. Smoothing averages the position of the segments which are part of the filaments over a smoothing length, but does not modify the position of the critical points nor the number of segments per filament.  Varying the smoothing strength has been tested and does not significantly affect the results (see Appendix~\ref{app:qualassessangle}).

Figure~\ref{fig:method} presents a zoom into the cosmic web reconstruction in the EDF-F, with a few galaxy cutouts to present the method. The closer the different realisations of a cosmic filament are, the more likely this filament is a real projected filament. Note that \cite{Q1-SP005} presented a match of the cosmic filaments extracted in a similar way with external cluster catalogues, which is another way to assess the robustness of the extraction.  For consistency, a quality assessment of the reconstruction is also provided in Appendix~\ref{sec:qualassess}, which presents results on the DAWN catalogues for comparison. 

In Appendix~\ref{sec:uncertain}, we present an estimate of the typical confusion on the filament positions based on the multiple realisations of the cosmic web in the same slice. We note little variation as a function of redshift and across fields. The median uncertainties $\Delta_{\rm skl}$ are
$0.76$, $0.68$, and  $ 0.70 \,h^{-1}{\rm Mpc}$ at $0.5<z<0.7$ and $ 0.85$, $ 0.80$, and $0.85\,h^{-1}{\rm Mpc}$ at $0.5<z<0.9$ in the EDF-N, EDF-S, and EDF-F respectively. 

\begin{figure*}
    \centering
   \includegraphics[width=0.99\linewidth]{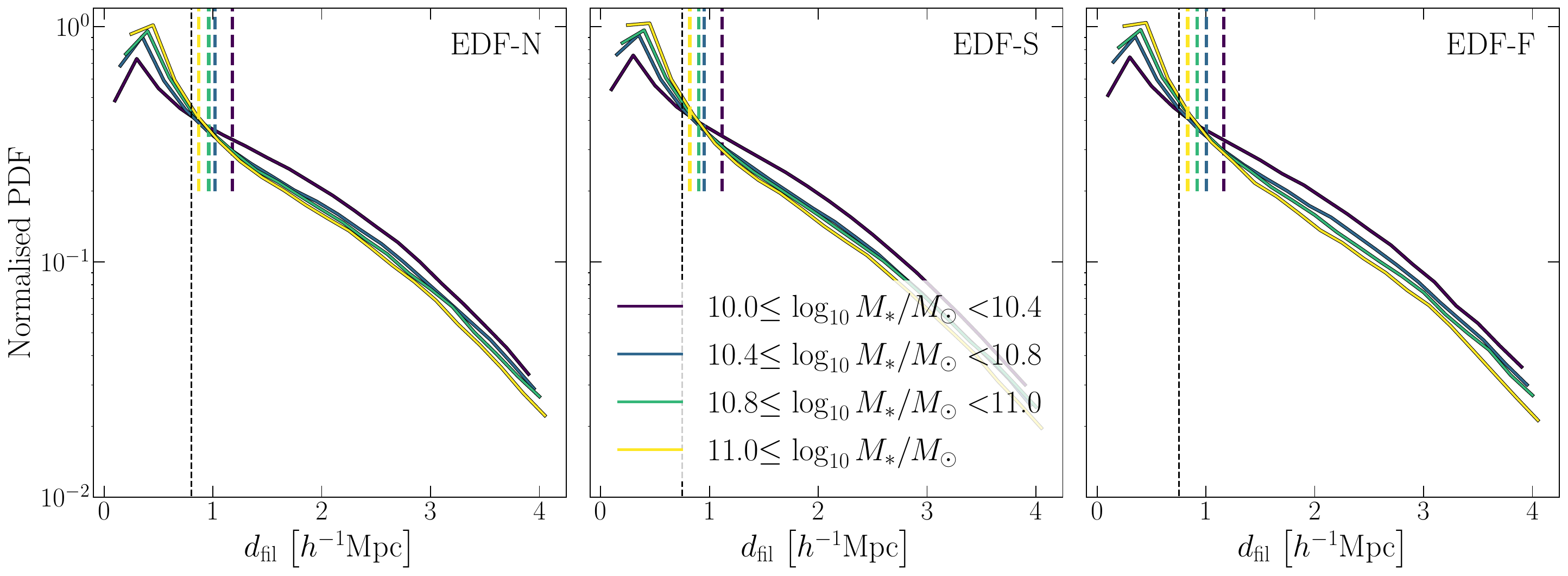}
    \caption{Distribution of galaxies around filaments in bins of masses for each of the three EDFs separately. The black vertical dashed line indicates the typical uncertainty on filament position as estimated in Fig.~\ref{fig:filwidth}. The coloured vertical dashed lines indicate the median of each distribution.}
    \label{fig:masses}
\end{figure*}

\begin{figure*}
    \centering
    \includegraphics[width=0.95\textwidth]{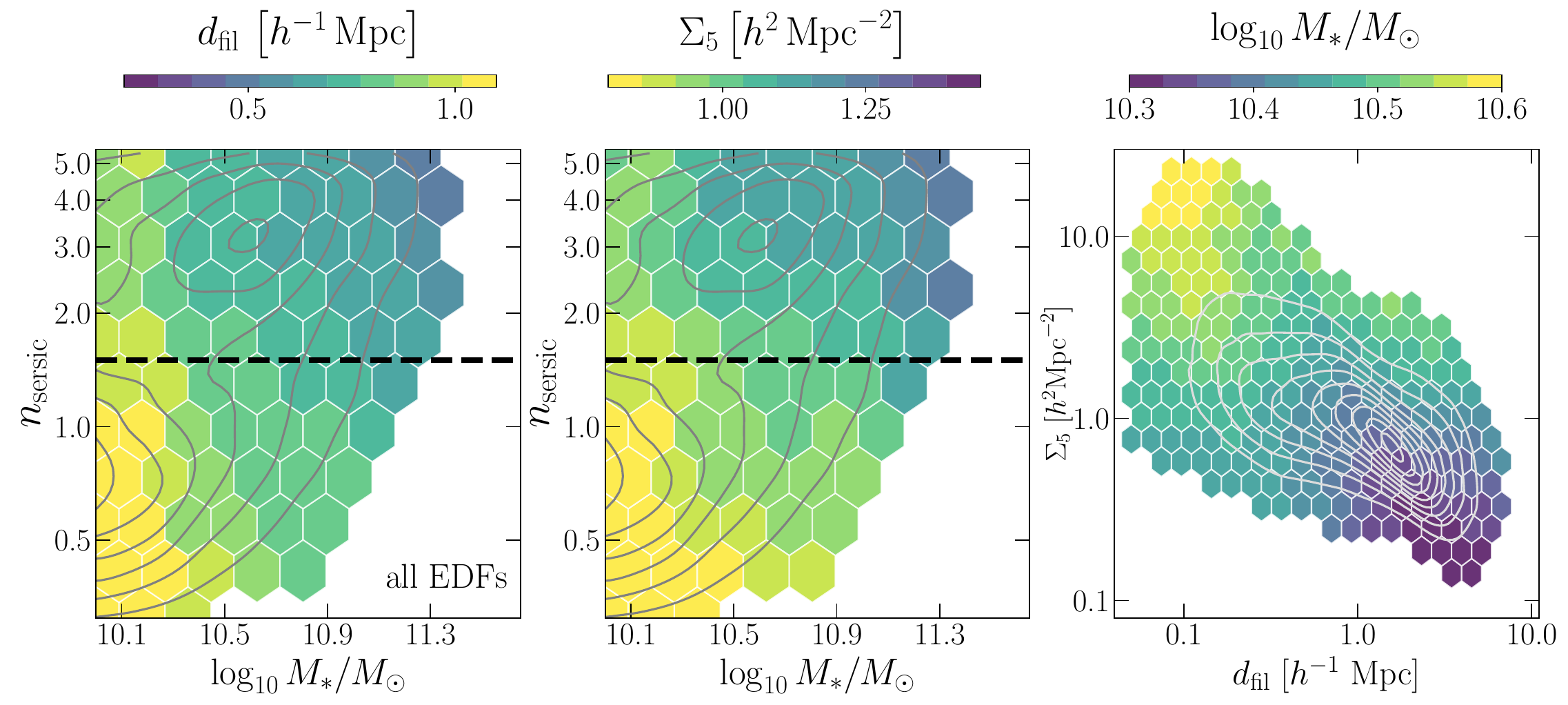}
    \caption{S\'ersic index versus mass distribution for galaxies in the selected-sample, colour-coded by median distance to filaments (left) and median density $\Sigma_5$ (middle) in all EDFs combined. Only hexagons with more than 200~galaxies are coloured.  Isocontours in grey are derived from Fig.~\ref{fig:bimodality}. The dashed black line represents the boundary of the early-type population, as defined in Sect.~\ref{sec:finsamp}. The right panel displays the correlation between density and distance to filaments, colour-coded by the median stellar mass in each hexagon. In this panel, grey lines are isocontours from the underlying galaxy distribution.}
    \label{fig:morpho_2}
\end{figure*}
\subsubsection{Density}
Finally, we also estimate the local density $\Sigma_N$, following \cite{baldry06}, from the distance to the  $N$th closest galaxy $d_N$ \citep[see also e.g,][]{bluck20,McDonough2025}. In each realisation of the slice, we compute
\begin{equation}
    \Sigma_N = \frac{N}{\pi d_N^2}\,,
\end{equation}
using all the galaxies with a stellar mass $\log_{10} (M_*/M_\odot) > 10.3$ in the 170~$h^{-1}{\rm Mpc}$ thick slices. We principally explored $\Sigma_5$ and $\Sigma_{10}$, but the results presented in Fig.~\ref{fig:morpho_2} are not strongly dependent on $N$, so we only displayed the measurement for $\Sigma_5$. Typically,  $\Sigma_5$ varies from 0.1~($h^{-1}{\rm Mpc}$)$^{-2}$ (low density) to 250~($h^{-1}{\rm Mpc}$)$^{-2}$ (high density).

\subsection{Measuring distances to filaments and galaxy orientation}

Galaxy isophotal position angles are measured with the \texttt{SourceXtractor++} software \citep{Bertin-2020-SourceXtractor-plus-plus, Kummel-2022-SE-use} from the VIS images by the MER pipeline \citep{Q1-TP004}. An elliptically symmetric 2-dimensional S\'ersic profile is fitted from the VIS images  \citep[see e.g.,][for a quality assessment of the angles]{Q1-SP040}. The fitting procedure takes the PSF model as an input file and the S\'ersic profile is convolved with the PSF prior to the fit, which minimises the contamination of position angles by the PSF. In addition, since the galaxies considered in this study are particularly well resolved, spurious alignment signal driven by PSF ellipticity is unlikely, which Appendix~\ref{app:qualassessangle} confirms through additional validation tests. Finally, we do not detect any signal when reshuffling galaxy position angles with respect to filaments (see Sect.~\ref{sec:alignment}), which further confirms the absence of spurious signal driven by systematics in the position angles themselves. This reshuffling was performed independently for early-type and late-type galaxies in order to make sure that there are no  morphology-dependent systematic effect in the galaxy position angles.

Only galaxies with their median {\tt NNPZ} falling in the smallest $z$-interval encompassing 70\% of the {\tt Phosphorus} PDF($z$) are kept in the sample. For the measurements presented below, we rely jointly on galaxy {\tt NNPZ} masses and redshifts. Because the joint probability of masses and redshifts are not released as part of the Q1 sample,  we cannot perform a fully probabilistic assignment of galaxies to tomographic slices. Therefore, we only associate galaxies with their most likely tomographic slice based on the mode of their {\tt Phosphorus} redshift {\tt phz\_mode\_1}. 
For each realisation of the filament extraction, the  distance to the closest filament and the position angles between the galaxy major axis and the closest filament segment are measured. The distance to the closest filament is then converted into comoving $h^{-1}{\rm Mpc}$ using the mean redshift of the tomographic slice. In the following, we call $\theta_{\rm fil}$ the position angle on the sky between the galaxy major axis and the closest filament, and $d_{\rm fil}$ the distance of the galaxy to its closest filament. Through this approach, we build for each galaxy a distribution of distances and angles with respect to cosmic filaments, as displayed in the sub-panels of Fig.~\ref{fig:method}.
Galaxy shapes and filaments are non-oriented angles and therefore the angles vary from~0$^{\circ}$ to~90$^{\circ}$. The expected averaged angle in the case of random alignment is therefore 45$^{\circ}$. 

\begin{figure}
    \centering
    \includegraphics[width=0.95\columnwidth]{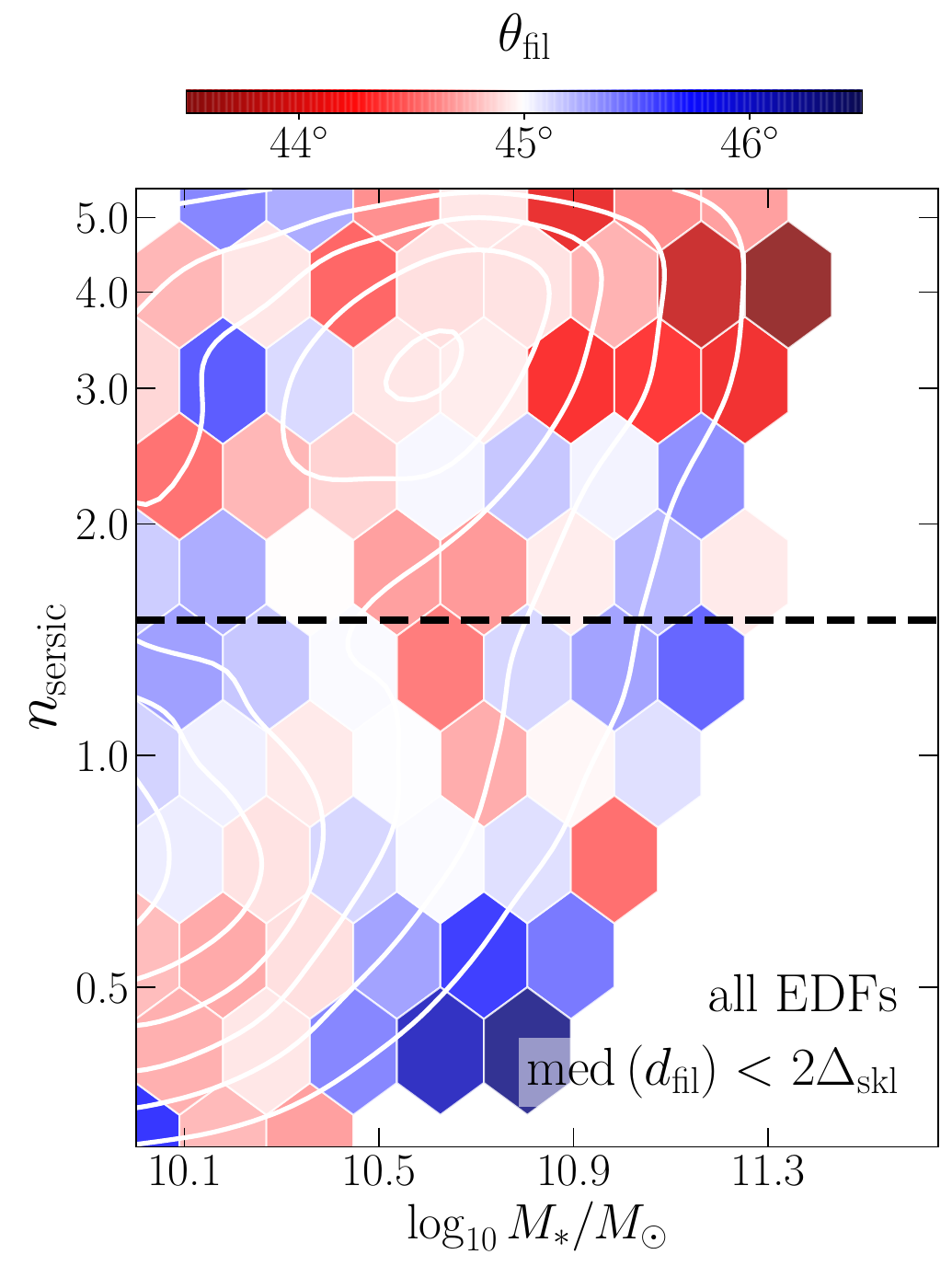}
    \caption{S\'ersic index versus mass distribution for galaxies in the selected-sample, colour-coded by the median angle between galaxy major axes and the closest filament $\theta_{\rm fil}$. All three fields are combined here and only galaxies with a median distance to filament smaller than 2$\Delta_{\rm skl}$ are kept. $\Delta_{\rm skl}$ is the mean uncertainty on filament position (see Sect.~\ref{sec:tool}).
  }  
    \label{fig:2dalign}
\end{figure}

\section{Results}
\label{sec:result}

\subsection{Statistics of galaxies in the cosmic web}
In order to check the homogeneity of the measurement quality  across the three EDFs, we present in Fig.~\ref{fig:masses} the distribution of galaxies around filaments in bins of mass for each of the three EDFs. For each galaxy, we compute the distance to the closest filament in all realisations of the skeleton, building a PDF of distances to filament for each galaxy. To build Fig.~\ref{fig:masses}, the PDF of distances to filaments for all galaxies were stacked together. There is no significant evolution of these distributions when splitting into different redshift bins. We also indicate with a black vertical dashed line the typical uncertainty on filament position (see Appendix~\ref{sec:uncertain}).  Unsurprisingly, more massive galaxies lie closer to the filament spines, as was highlighted in many previous works \citep[e.g.,][]{Malavasi17,Kraljic18,Laigle2018,Kane2024}. This effect is at first order driven by the fact that filaments are large-scale overdensities. They induce
a density boost, allowing the proto-halo to pass the critical threshold of collapse earlier \citep{Kaiser1984}. Besides the basic effect of the amplitude of the local density, the cosmic web  imposes tides, which will modulate the halo accretion rate as a function of distance to saddles and nodes of the filaments \citep{musso18}. As a consequence of these two effects, the halo mass function is expected to be biased in the vicinity of the large-scale cosmic filaments \citep[e.g.,][]{Alonso15}. Since galaxy mass is mainly driven by halo mass, the natural consequence is to find more massive galaxies closer to the filament spines, which is also noticeable from the large-scale clustering of galaxies as a function of mass \citep[e.g.,][]{alam2019}.

\subsection{Galaxy morphologies in the cosmic web}

Figure~\ref{fig:morpho_2} presents the distribution of  S\'ersic index versus mass for galaxies in the selected sample, colour-coded by mean distance of galaxies to filaments and local density, $\Sigma_5$, measured from the inverse distance to the 5th neighbour. To build this figure, the distributions of distances and densities obtained for each galaxy from the different slice realisations are stacked. Each hexagon is coloured by the median of the stacked distribution in the hexagon. The measurements of all EDFs are combined here. The median distance to filaments smoothly varies across this diagram as a function of both stellar mass and S\'ersic index, early type galaxies being on average closer to filaments than late-type galaxies. This is still true at fixed stellar mass: at a given stellar mass, galaxies that have a late-type morphology are more likely to lie further away from the cosmic filament spines. 
We note that the gradients in distance to filaments and density are similar, but not completely equivalent, suggesting that distance to filaments and density are not completely interchangeable. This is further illustrated on the right panel of this figure, which displays the correlation between density and distance to filament, colour-coded by the median stellar mass of galaxies falling into each hexagon. Stellar mass gradients are neither horizontal nor vertical, suggesting that both environmental estimators play a role in shaping mass assembly. 

\begin{figure*}
\centering\includegraphics[width=0.99\textwidth]{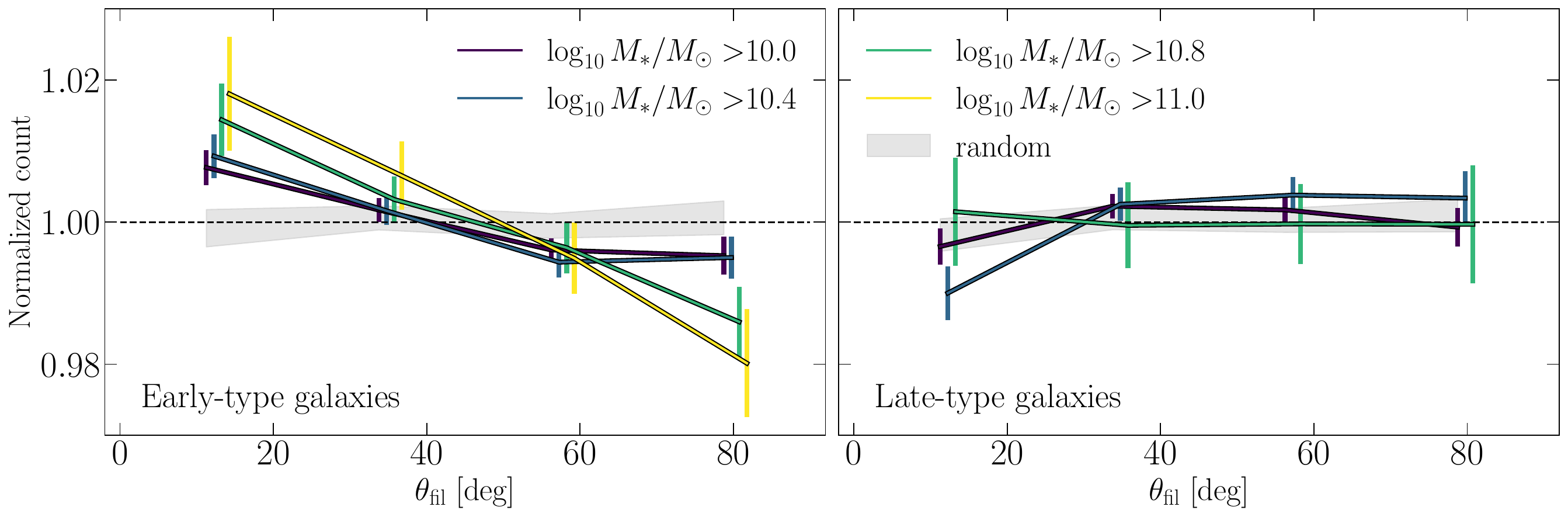}
\caption{Normalised distribution of angles for early-type galaxies (left), and late-type galaxies (right) in the three EDFs, for different mass thresholds. The grey area corresponds to the random signal, which is obtained by reshuffling galaxy orientations while keeping fixed their other properties (especially positions with respect to filaments). The numbers of galaxies above the four mass thresholds, are, by increasing mass: 47229, 35134, 13827, and 5867 for the early-type galaxies and 45933, 18549, 4133, and 1482 for the late-type galaxies. The distribution above a given mass threshold is displayed only if it is computed with more than 2000 galaxies.}
\label{fig:alignment}
\end{figure*}

\subsection{Shape alignment with cosmic filaments}
\label{sec:alignment}
Figure~\ref{fig:2dalign} presents the 2D distribution of galaxies in the plane defined by S\'ersic index and masses, but colour-coded by their angle between galaxy major axis and their closest filaments ($\theta_{\rm fil}$). To build this figure, the distributions of distances and densities obtained for each galaxy from the different slice realisations are stacked. In addition, to account for uncertainties on galaxy masses and S\'ersic indices, 16 realisations of masses and S\'ersic indices are drawn for each galaxy by assuming for each quantity (either mass or S\'ersic index) a normal distribution with a standard deviation matching its $1\,\sigma$-uncertainty. Each hexagon is coloured by the median of the stacked distributions of galaxies falling in the hexagon. Figure~\ref{fig:2dalign}  presents the result after keeping only galaxies with a median distance to filament smaller than twice $\Delta_{\rm skl}$ (where $\Delta_{\rm skl}$ is the typical uncertainty on filament positions, which is of the order of $0.8$~$h^{-1}$Mpc, see Sect.~\ref{sec:tool} and Fig.~\ref{fig:filwidth}). For early-type galaxies, we notice by eye a trend for $\theta_{\rm fil}$ to systematically depart from random orientation (which would be characterised by $\langle \theta_{\rm fil}\rangle \simeq 45^{\circ}$ since the angles are in  the range $0^{\circ}$--$90^{\circ}$). A weak signal  for late-type galaxies to have $\theta_{\rm fil}$ smaller than 45$^{\circ}$ seems noticeable.  In Fig.~\ref{fig:standerror}, we present the standard error on the mean in each hexagon.

To further quantify this trend, we then separate the population of early-type and late-type galaxies, using the boundary line defined in Sect.~\ref{sec:finsamp}.  
The stacked normalised distribution of angles for the early-type population on the one hand, and for the rest of the population on the other hand, is displayed in Fig.~\ref{fig:alignment}, as a function of stellar mass threshold. The grey area corresponds to random position angles, obtained by reshuffling galaxy position angles (across each field independently) within a population (early-type or late-type above $\log_{10} (M/M_\odot)>10$) while keeping fixed the rest of their properties (position and mass). To build this figure, we add up the contribution of all angle measurements in all realisations, so as to account for the spread of the angle probability distribution function in the measurement. Error bars are obtained by bootstrapping the distribution of galaxies 100~times.
The alignment signal between galaxy major axes and cosmic web filaments increases as we isolate more massive galaxies. Given the error bars, the signal is significant at all masses. No significant perpendicular alignment is detected for late-type galaxies given the error bars, except at the stellar mass threshold $\log_{\rm 10}(M_*/M_\odot)>10.4$. We note however that our cut on stellar mass does not allow us to explore the perpendicular alignment signal of late-type galaxies at low-mass, where it is potentially stronger. In addition, the cut at $n_{\rm sersic}=1.5$ to distinguish between early-type and late-type galaxies might be too coarse and not optimal. We also expect the alignment signal to be much stronger for early-type galaxies, especially at moderate redshift \citep[see][on the {Horizon-AGN} simulation]{Codis2018}. 
In Appendix~\ref{app:qualassessangle}, we present the result for the three different fields independently, with different selection cuts, variations of the skeleton extraction parametrisation, and two methods to estimate the uncertainties (bootstrap or jackknife resampling). The alignment signal of early-type galaxies remains robust to all these changes, as well as the perpendicular alignment signal of late-type galaxies with stellar mass $\log_{\rm 10}(M_*/M_\odot)>10.4$ (although no signal is found when isolating the most massive galaxies in this mass range). We also note that increasing the $n_{\rm sersic}$ threshold to select early-type galaxies strengthen their alignment signal.

\section{Discussion}
\label{sec:conclusion}

Using the Q1 data set, we have investigated the correlation between galaxy shapes and their surrounding cosmic web filaments. We have extracted the distribution of projected cosmic filaments in $170\,h^{-1}{\rm Mpc}$-thick tomographic slices over the redshift range $0.5<z<0.9$, using about 100~000 galaxies more massive than $10^{10}M_\odot$  in our more conservative sample. 
The main findings of this study are as follows:
\begin{itemize}
    \item We  recover projected cosmic filaments (Fig.~\ref{fig:method}), with a positional uncertainty driven by photometric redshifts of the order of $\simeq 0.8$~$h^{-1}{\rm Mpc}$. The comparison to external data sets supports the reliability of the extraction (Appendix~\ref{sec:qualassess}).
    \item We find in each of the three EDFs that massive galaxies are more clustered around projected filament spines (Fig.~\ref{fig:masses}). To first order, this can be attributed to the biased halo mass function within filaments, driven by the increased probability of collapse in overdense regions on the one hand, and to the modulation of halo accretion rates (anisotropic assembly bias) influenced by the tidal forces exerted by the cosmic web on the other hand. At this stage, we cannot distinguish between the different possible causes for this trend and we specifically cannot claim evidences for assembly bias.
    \item The average distances of galaxies to filaments vary with their S\'ersic indices and masses (Fig.~\ref{fig:morpho_2}). This trend is not entirely equivalent to changes in local density. Early-type galaxies are observed to lie closer to the filament spines than late-type galaxies, suggesting that they are more prone to environmental interactions within cosmic filaments, which can drive morphological transformations.
    \item When analysing the orientation between the projected major axis of galaxies and the direction of projected filaments for increasing mass thresholds, we observe an increased probability for early-type galaxies to align with the projected filaments (Fig.~\ref{fig:alignment}). There is a weak signal for late-type galaxies with $\log_{10} (M_*/M_\odot) > 10.4$ to depart from random orientations (in the sense of a perpendicular alignment).
\end{itemize}

This work confirms the trend that early-type galaxy major axes tend to align with cosmic web filaments in a mass-dependent fashion, a result which was already highlighted in previous results, but at lower redshifts \citep[e.g.,][]{Singh2015,Chen2019,DesaiRyden2022}. Alignment of late-type and low-mass galaxies, if it exists, is expected to be more subtle \citep[see e.g.,][]{Codis2018}. The lack of significant detection in our work except at $\log_{10} (M_*/M_\odot) > 10.4$ could be the consequence of projection effects due to the large slice thickness, as well as the photometric redshift and stellar mass uncertainties, or a combination of both. A proper investigation of the alignment of low-mass galaxies still needs to be carried out, since we restricted  our measurements here to $\log_{10}(M_*/M_\odot)>10$. Furthermore, the separation between early-type and late-type galaxies could be refined by optimising the cut on the S\'ersic index.

In addition, we note that the effect of projection, especially in thick slices, has inevitably a non-negligible impact on the cosmic web reconstruction and on the alignment signal. In thick slices, filaments cannot be disentangled from matter sheets seen in projection, and nodes, which in 3D are maxima of the density field in the direction towards which filaments are converging, can simply in 2D be overlapping projected filaments, that are not necessarily connected in the 3D volume. Moreover, physically unrelated peaks of the density field in 3D may appear close in 2D, which can drive the detection of filaments linking them. As for the alignment signal with filaments, it can also be strongly reduced depending on the orientation of the filament with respect to the viewing angle \citep{Zhang2023}. Such blurring effects of the signal could be at least partially overcome with the large statistics that the EWS will offer us.
Additional analyses on simulations should be carried out to further optimise the cosmic web extraction method and its parametrisation. 

Although this study was conducted using the Q1 data set -- a very small subset of the final Euclid survey -- it already highlights our future ability to precisely characterise the cosmic web environment of galaxies with the future EDS. 
Future Euclid releases will enhance these measurements further, extending the redshift and mass ranges explored in this study. This will be made possible thanks to the excellent quality of the photometric redshifts that is expected in the EDFs when they will reach their full depth and will be combined with data from the DAWN survey. Reducing the uncertainties on photometric redshifts will also allow us to reduce the thickness of the tomographic slices and therefore to minimise the associated projection effects, hence enhancing the quality of the cosmic web reconstruction. We can therefore expect that we will be able to conduct finer analyses of the impact of the cosmic web environment on galaxy properties including their morphologies \citep[see e.g.,][]{Q1-SP040} and star-formation state \citep[see e.g.,][]{Q1-SP031, Q1-SP044}, on an unparalleled large survey area. In particular, future work will analyse these trends not only in terms of stellar masses and S\'ersic indices, but also as a function of galaxy bulge existence and prominence, which are parameters related to the galaxy quenching state \citep{lang14,bremer18,bluck20,bluck22,dimauro22,quilley22}. These parameters can be quantified based on the excellent photometry from VIS \citep{EuclidSkyVIS,Q1-TP002} and powerful tools developed by the Euclid Consortium to characterise galaxy morphologies \citep{Q1-TP004,Q1-SP047}. Finally, when the Euclid spectroscopic sample produced thanks to the NISP instrument \citep{EuclidSkyNISP} will reach it full depth in the EDFs, we anticipate that the 3D-reconstruction of the cosmic network will be possible (Euclid Collaboration: Kraljic et al., in prep.). This will open new possibilities for analysing correlations between galaxy orientations and the cosmic web without suffering projection effects.

\begin{acknowledgements} 
\AckQone 
\AckEC 
\AckDatalabs 
This work makes use of the DR1 catalogues from the Cosmic Dawn Survey delivered to the Euclid consortium.
This work is also based on data from UNIONS, a scientific collaboration using  
three Hawaii-based telescopes: CFHT, Pan-STARRS, and Subaru  
\url{[www.skysurvey.cc](http://www.skysurvey.cc)} and on data from the Dark Energy Camera (DECam) on the Blanco 4-m Telescope  
at CTIO in Chile \url{[https://www.darkenergysurvey.org](https://www.darkenergysurvey.org)}.
 
CP is supported by the French Agence Nationale de la Recherche under Grant \href{https://www.secular-evolution.org}{SEGAL}  ANR-19-CE31-0017.
We thank St\'ephane Rouberol for the smooth running of the Infinity cluster, where part 
of the analysis was performed.
\end{acknowledgements}

\twocolumn
\clearpage

\bibliography{biblio,Q1,Euclid}


\newpage

\begin{appendix}
\onecolumn

\section{Properties of the fiducial sample}

\begin{figure*}
    \centering
    \includegraphics[width=0.95\linewidth]{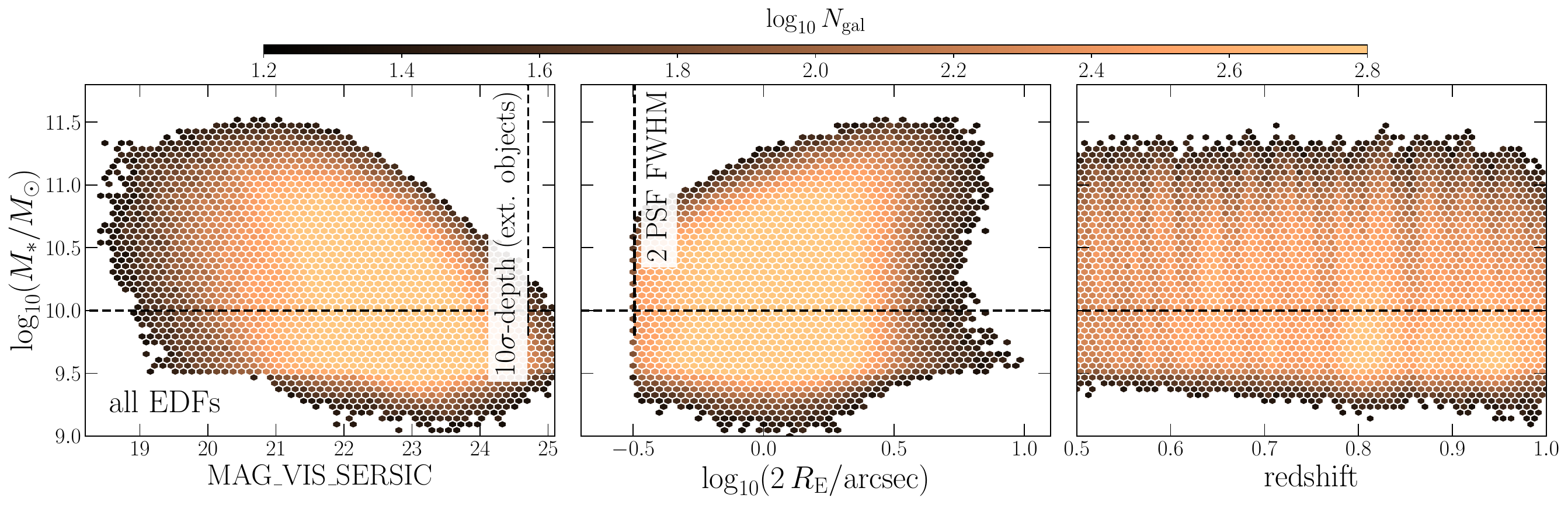}
    \caption{Properties of our fiducial sample. \textit{Left}: Distribution of stellar mass versus VIS magnitude. The vertical dashed line indicates the depth at $10\,\sigma$ for extended objects: all galaxies above our stellar mass limit of $10^{10} M_\odot$ have a signal-to-noise ratio larger than 10. \textit{Middle}: Distribution of stellar mass versus the effective diameter estimated from the S\'ersic fit. The vertical lines indicate twice the PSF FWHM. By selection, all galaxies in our fiducial sample have a diameter larger than twice the PSF FWHM, so they can be considered as resolved. \textit{Right}: Stellar mass versus redshift. The stellar mass cut at $10^{10} M_\odot$ ensures a complete sample in the redshift range $0.5<z<1$.}
    \label{fig:propsample}
\end{figure*}

\begin{figure*}
    \centering
    \includegraphics[width=0.94\textwidth]{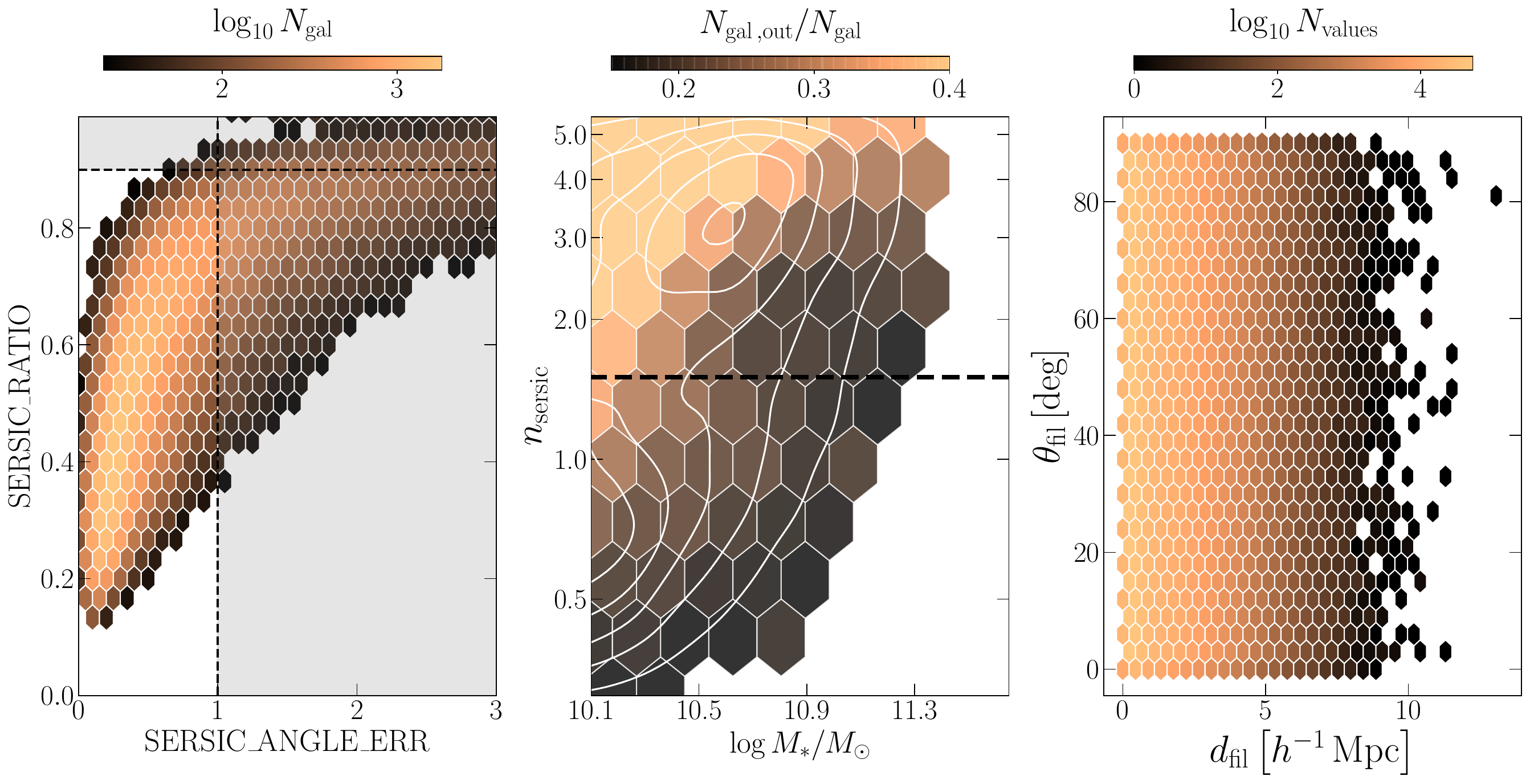}
    \caption{Properties of our fiducial sample. \textit{Left}: Distribution of S\'ersic ratio and uncertainties on the position angle. Our fiducial sample contains only galaxies with a  S\'ersic ratio smaller than 0.9 and an error on the position angle smaller then one. The grey transparent area indicates galaxies that are rejected by our selection. Their properties are displayed in the middle and right panels. \textit{Middle:} Fraction of rejected galaxies $N_{\rm gal\, out}$/$N_{\rm gal}$ as a function of S\'ersic index and stellar masses. \textit{Right}: Distribution of rejected galaxies in terms of distances to filaments $d_{\rm fil}$ and alignment angles $\theta_{\rm fil}$.}
    \label{fig:prop_selection}
\end{figure*}

Figure~\ref{fig:propsample} presents the properties of our fiducial sample, in comparison to the depth and FWHM of the VIS images. More specifically, the figure highlights that our selected galaxies have a signal-to-noise ratio larger than 10 (left panel), are resolved (their diameters are larger than twice the VIS PSF FWHM, middle panel) and our cut at $\log_{10}(M_*/M_\odot) > 10$ for the main sample analysed in this work is well above the stellar mass completeness.

To further characterise the fiducial catalogue, we quantify how the cut at 0.9 in {\tt SERSIC\_RATIO} and at~1 in {\tt SERSIC\_ANGLE\_ERR} potentially biases the sample. Figure~\ref{fig:prop_selection} shows the selection on the left panel, the S\'ersic index and mass distribution of the rejected galaxies (those which fall on the shaded area in the left panel) in the middle panel, and the distribution of $\theta_{\rm fil}$ and $d_{\rm fil}$ on the right panel. Galaxies with {\tt SERSIC\_RATIO} close to~1 and large {\tt SERSIC\_ANGLE\_ERR} are distributed everywhere, but they are more numerous at very high S\'ersic indexes and low masses. They do not occupy specific locations in the diagram of $\theta_{\rm fil}$ versus $d_{\rm fil}$. However excluding them or not does not significantly change the alignment signal (see Fig.~\ref{fig:variations_selection}). 

\section{Quality assessment of the cosmic web reconstruction}
\label{sec:uncertain}
\label{sec:qualassess}

\subsection{Uncertainties on filament positions}

\begin{figure*}
    \centering
    \includegraphics[width=0.95\linewidth]{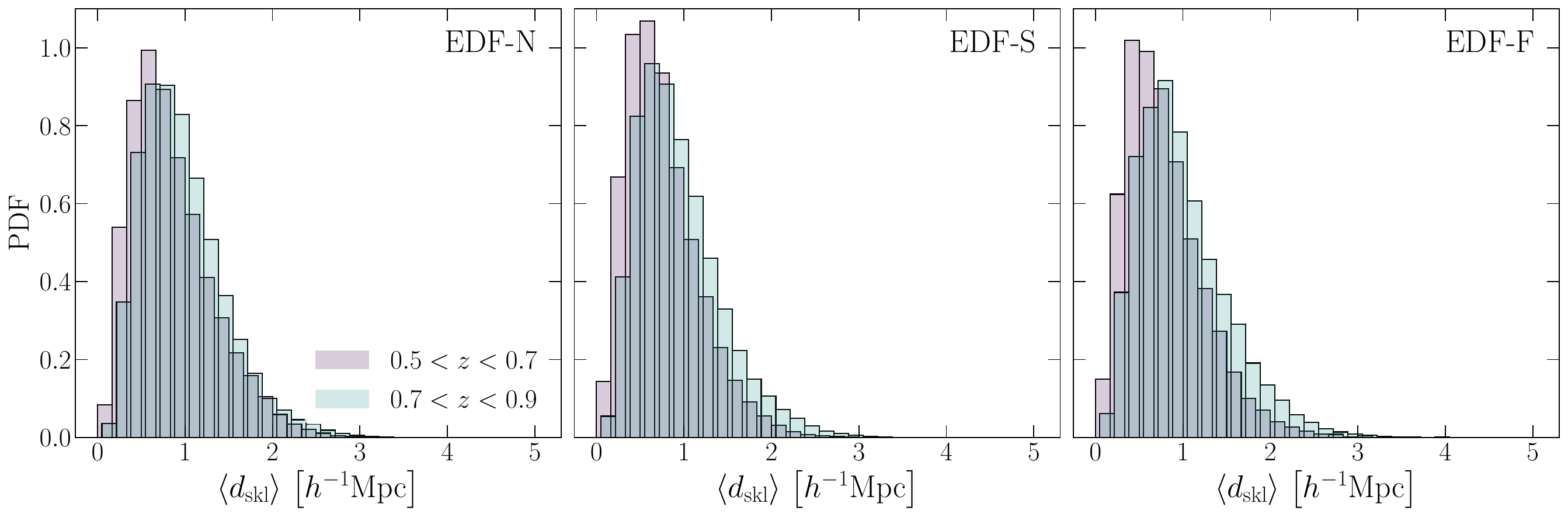}
    \caption{Typical confusion on filament positions due to the photometric redshift uncertainties. For a given segment in a given tomographic slice, we measure the mean of the distances to the closest counterparts in all  other realisations of the skeleton.}
    \label{fig:filwidth}
\end{figure*}

Figure~\ref{fig:filwidth} presents an estimate of the typical confusion on the filament positions. For a given segment, we measure the mean of the distances to the closest counterparts in all other realisations of the cosmic web in the same tomographic slice. We note little variation as a function of redshift, but the confusion is slightly higher at low redshift in the EDF-N than in the two other fields. The median of these distributions are
$0.76$, $0.68$ and  $ 0.70\,h^{-1}{\rm Mpc}$ at $0.5<z<0.7$ and $0.85$, $0.80$, and $0.85\,h^{-1}{\rm Mpc}$ at $0.5<z<0.9$ in the EDF-N, EDF-S, and EDF-F respectively.

\subsection{Comparison to the Cosmic Dawn Survey}

\begin{figure*}
    \centering
    \includegraphics[width=0.33\linewidth]{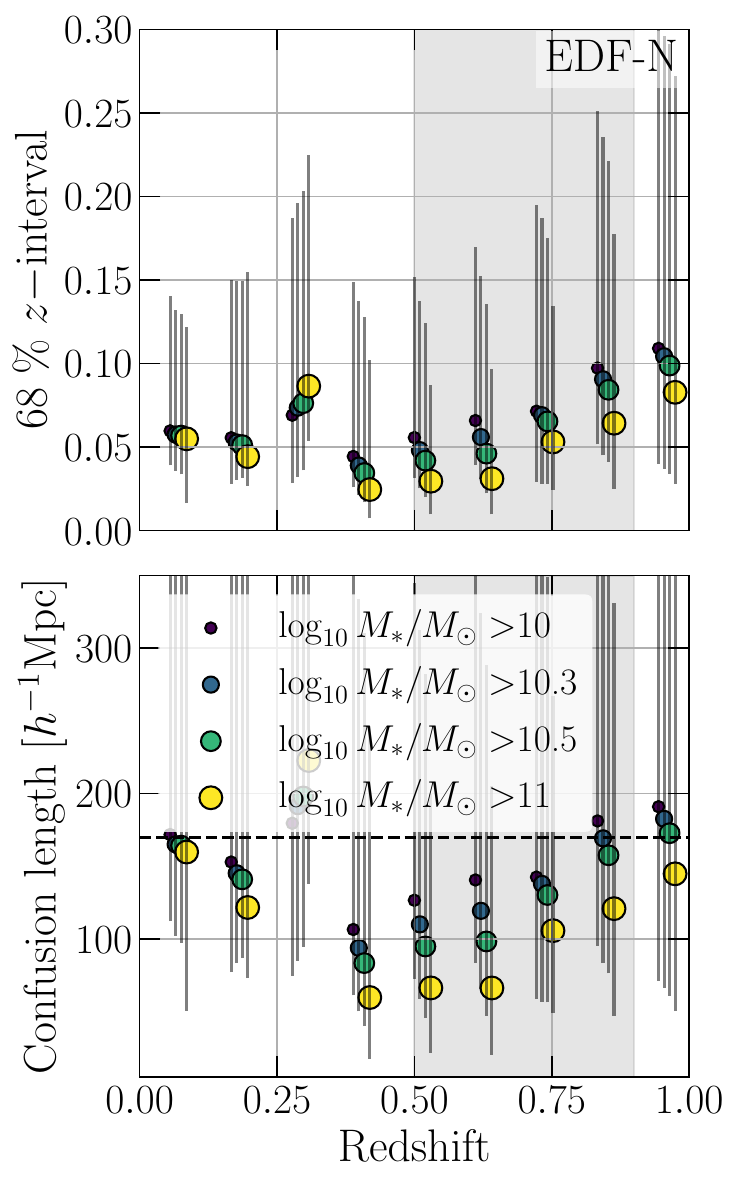}
    \includegraphics[width=0.33\linewidth]{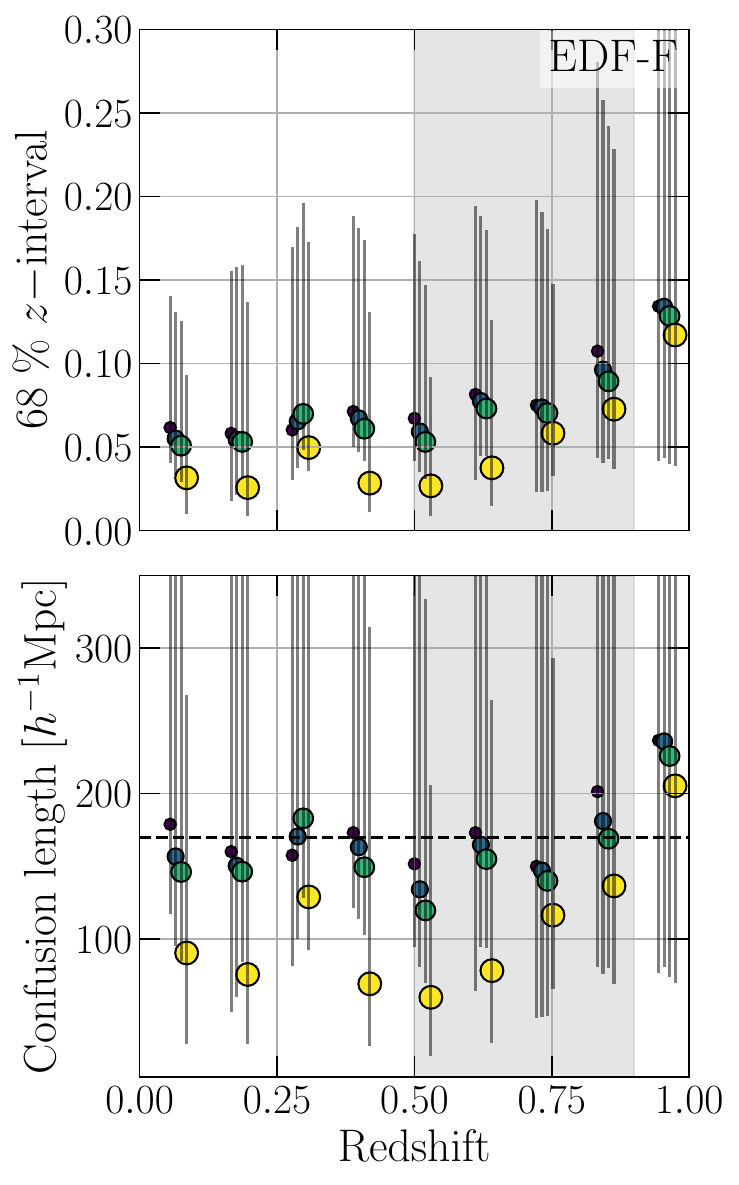}
    \caption{Similar to Fig.~\ref{fig:confusion_length}, but for galaxies in the DAWN catalogues. \textit{Top panels}: Redshift interval encompassing 68\% of the PDF($z$) around the median redshift (derived with \texttt{LePhare}) as a function of redshifts and stellar masses in the two fields covered by the DAWN survey. \textit{Bottom panels}: Corresponding uncertainty in $h^{-1}{\rm Mpc}$ on galaxy positions along the line of sight. }
    \label{fig:confusion_length_cd}
\end{figure*}

\begin{figure*}
    \centering
    
    \includegraphics[width=0.9\linewidth]{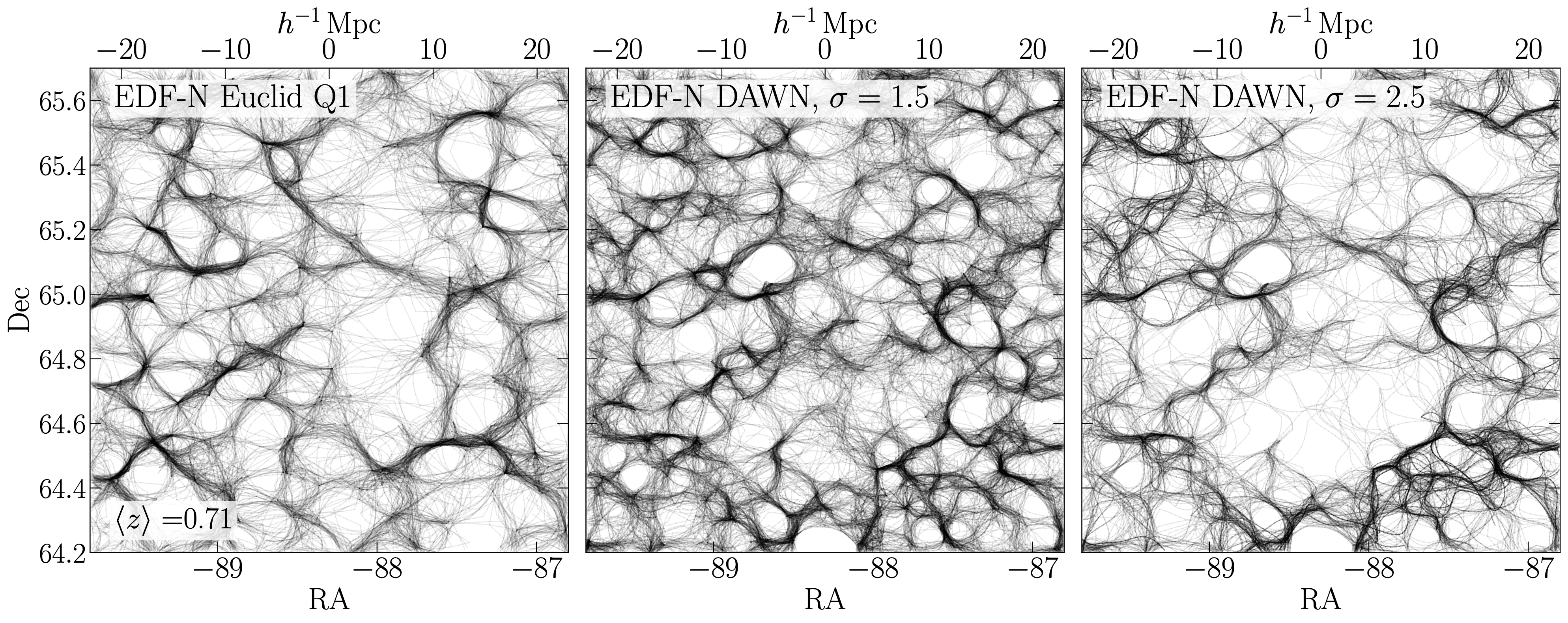}
    \includegraphics[width=0.9\linewidth]{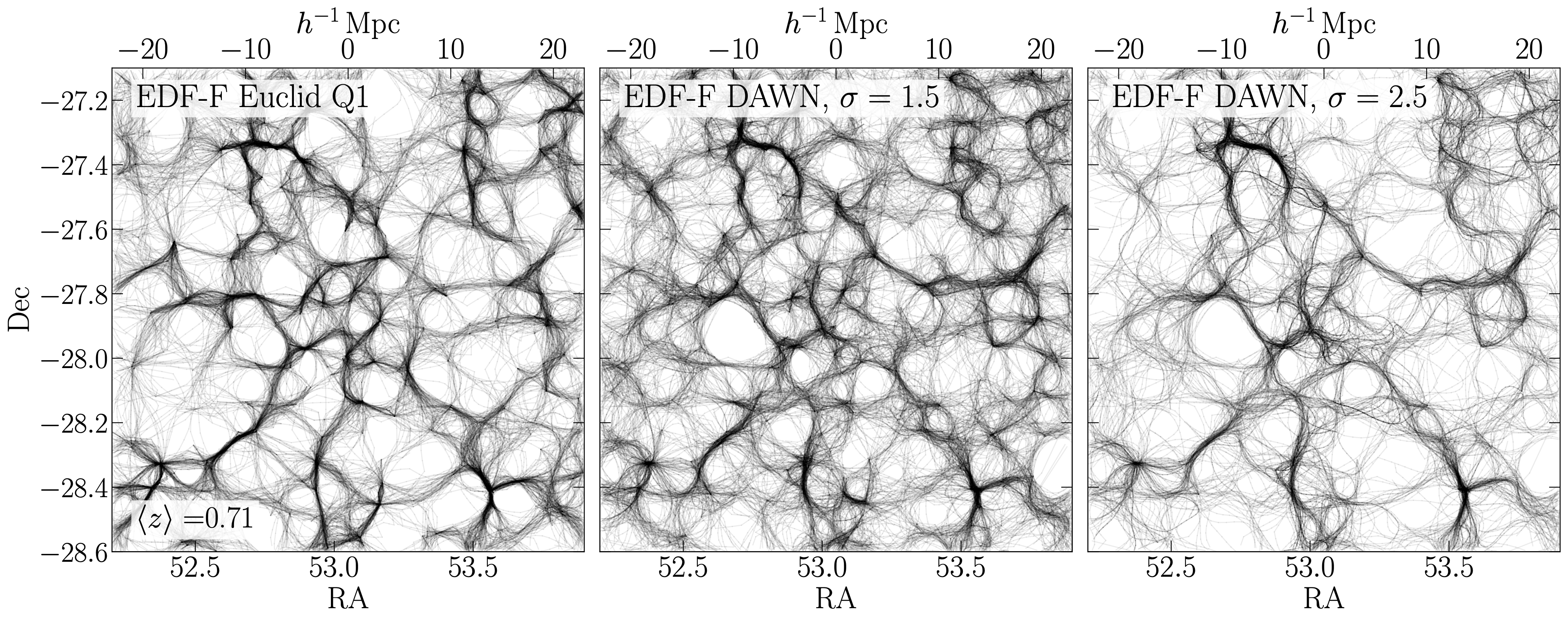}
    \caption{Visualisation of the cosmic web reconstruction based on the Q1 data set with a persistence threshold of $\sigma = 1.5$ (left) and the DAWN catalogues with a persistence threshold of $\sigma = 1.5$ (middle) and  $\sigma = 2.5$ at $z = 0.71$ in the EDF-N (top panels) and EDF-F (bottom panels).}
    \label{fig:visu_comparison}
\end{figure*}

 \begin{figure*}
     \centering
    \includegraphics[width=0.9\linewidth]{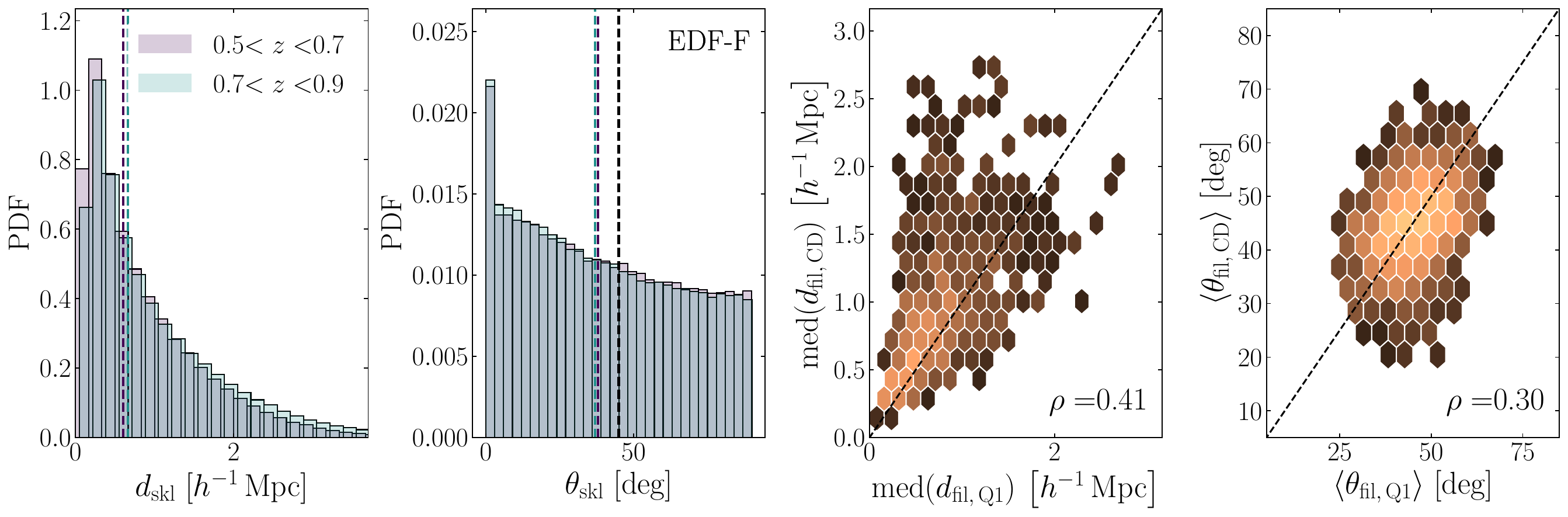}  \includegraphics[width=0.9\linewidth]{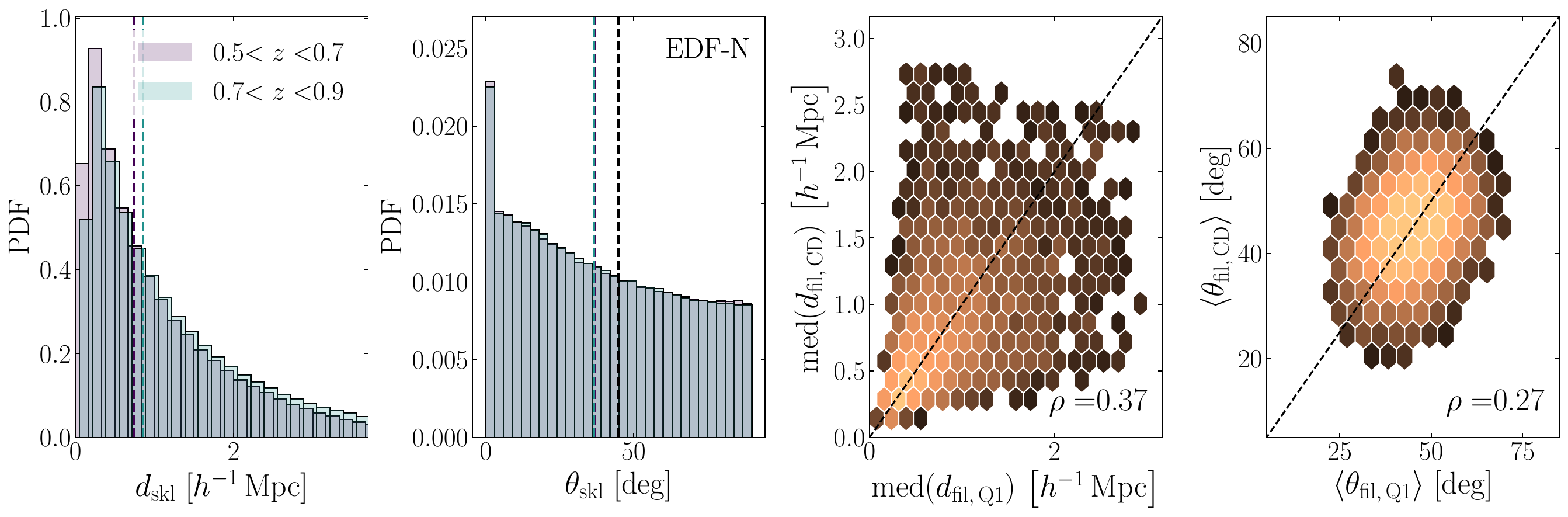}
     \caption{Quality assessment of the skeleton extracted in Q1, from the comparison with the DAWN skeletons in the EDF-F (top) and EDF-N (bottom). The left and middle left panels show the distribution of distances and angles between both skeleton sets. The middle right and right panel show how distances and angle of galaxies with respect to the DAWN skeleton compare with those measured with respect to the Q1 skeleton. The correlation coefficient is given in the bottom right corner of each subpanel. When related to the DAWN skeleton, galaxies from the Q1 release were positionally matched with the DAWN catalogues, and got assigned the redshift of the DAWN catalogue.}
     \label{fig:comparison_skeleton}
 \end{figure*}
We rely on the DAWN catalogues described in Sect.~\ref{sec:data} (covering part of the EDF-F and EDF-N) to perform a quality assessment of the cosmic web extraction, because the photometric redshifts are significantly better than in the Q1 release. In order to remove possible artefacts and filter only galaxies with reliable photometric redshifts, we apply the following cut:
\begin{itemize}
    \item \texttt{MODEL\_FLAG} $=0$, the \texttt{TRACTOR} model has converged;
    \item \texttt{SOLUTION\_MODEL} is a galaxy;
    \item \texttt{lp\_NbFilt} $>5$, more than five~photometric filters have been used to derive galaxy parameters;
    \item \texttt{HSC\_X\_VALID} the photometry is trustworthy in at least one of the HSC filters.
\end{itemize}

Figure~\ref{fig:confusion_length_cd} presents the photometric redshift uncertainties in this data set, estimated from the interval containing 68\% of the PDF($z$) around the median redshift. This estimator is comparable to the one presented in Fig.~\ref{fig:confusion_length}. In the EDF-N, the typical photometric redshift uncertainty at $z= 0.6$ is $ 0.04$ in the DAWN catalogues, while it is $0.08$ in the Q1 sample for galaxies with $\log_{10} (M_*/M_\odot) > 10.3$. In EDF-F, the improvement is noticeable with the DAWN catalogues, but less substantial. With much better photometric redshifts, the 2D cosmic web extracted in tomographic slices will be closer to the true projected cosmic structures and can be used as a reference. 
Figure~\ref{fig:visu_comparison} presents a visualisation of the tomographic slice at $z= 0.71$ in the DAWN and in the Q1 data sets, in both the EDF-F and the EDF-N. Visually, the agreement is good. In particular, we note that the cosmic filaments reconstructed in the Q1 data set have in general a counterpart in the DAWN filament data sets, but these filaments are not necessarily the most significant ones, even when comparing to a skeleton extracted with a higher persistence ($\sigma = 2.5$). 
Many filaments in the DAWN skeleton do not have a counterpart in the Q1 skeleton.

We further quantify the similarity between both cosmic-web filament data sets, following an approach outlined in Euclid Collaboration: Malavasi et al., in prep.
For each segment in the Q1 skeleton that overlap with the DAWN area, we look for its closer counterparts in the DAWN realisations of the skeleton. We then trace the distribution of the distances to DAWN filaments and the distribution of angles between both skeleton segments. We also compare the distance of galaxies to filaments in both skeletons, as well as their orientations with respect to filaments, for galaxies in our fiducial sample. When related to the DAWN skeleton, the Q1 galaxies were first matched with the DAWN catalogues based on their sky position, and we attribute to them the photometric redshift from the DAWN catalogue. The results are presented in Fig.~\ref{fig:comparison_skeleton}. We find that the median best-match distances are  $0.6$~$h^{-1}{\rm Mpc}$ and  $ 0.7$~$h^{-1}{\rm Mpc}$ at $0.5<z<0.7$ and $ 0.7$~$h^{-1}{\rm Mpc}$ and  $ 0.8$~$h^{-1}{\rm Mpc}$ at $0.7<z<0.9$ in EDF-F and EDF-N, respectively, while the median angles between skeleton sets are $38$~degrees and   $37$~degrees at $0.5<z<0.7$ in EDF-F and EDF-N, respectively. Remarkably, we note that although the quality of the reconstruction is poorer in Q1 data compared to DAWN in EDF-N, there is still a similar agreement (in terms of distance to skeleton and orientations) with the DAWN skeleton than in EDF-F. In other words, many DAWN filaments are missed in the EDF-N Q1 skeleton (as seen in Fig.~\ref{fig:visu_comparison}), but those filaments that are reconstructed are robust in the sense that they are present in the DAWN data set. Obviously, it should be acknowledged that the DAWN catalogues cannot replace a simulation to conduct dedicated tests on the quality of the extracted cosmic web skeleton.

Finally, we also compute the alignment of galaxy shapes with respect to the DAWN skeleton. The result is presented in Fig.~\ref{fig:variations_skeleton}.

\section{Quality assessment of the measurements of angles with respect to filaments}
\label{app:qualassessangle}

\subsection{Impact of the selection and skeleton parameterisation}
 \begin{figure*}
     \centering
    \includegraphics[width=0.95\linewidth]{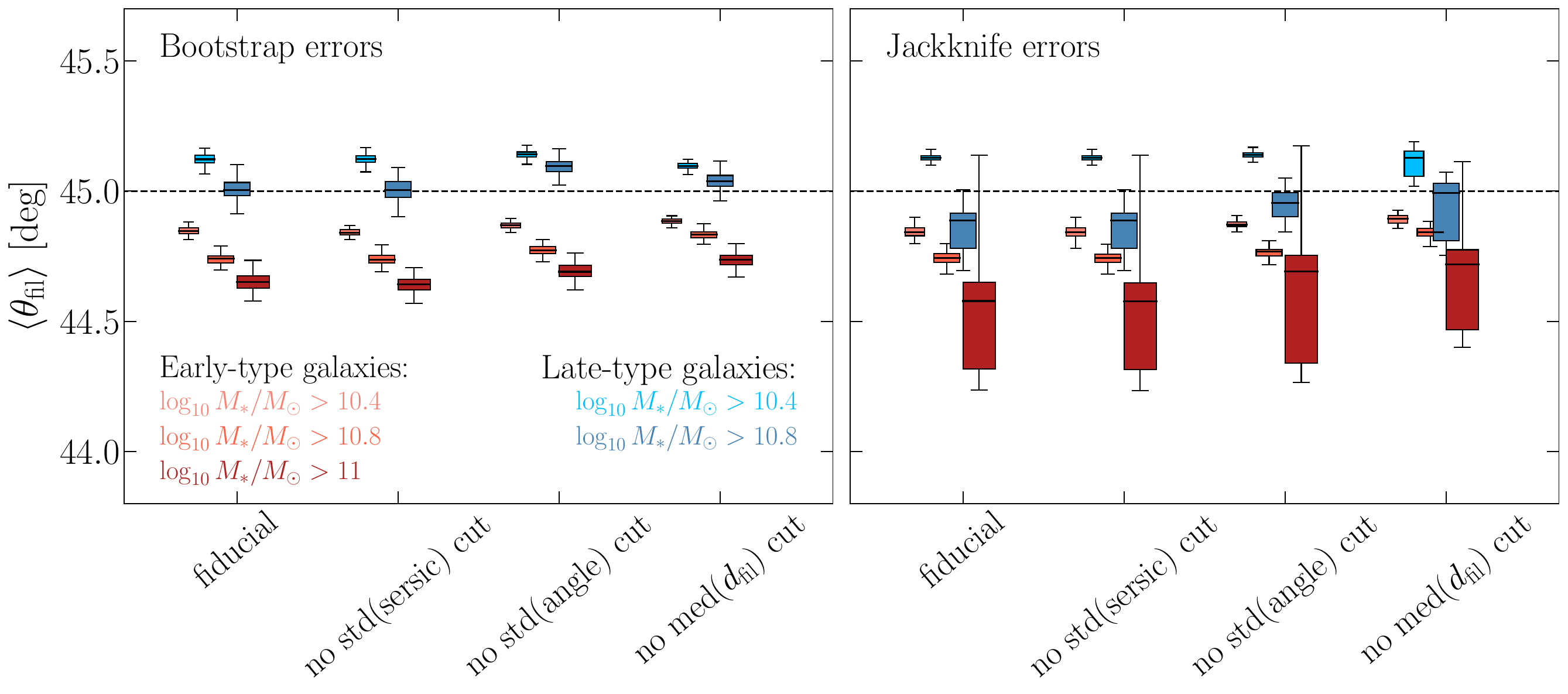}  
     \caption{Box plots displaying the alignment angle $\theta_{\rm fil}$ of early-type (red) and late-type (blue) galaxies at different mass thresholds for various selections of the sample, with uncertainties estimated from bootstrap (left) or jackknife (right) methods. While the cuts on {\tt Sersic\_index} (middle left), {\tt Sersic\_ratio} and {\tt Sersic\_angle\_err} (middle right) have little effect, relaxing the criterion on the median distance of galaxies to filament (right) reduces the signal for both late-type and early-type galaxies.}
     \label{fig:variations_selection}
 \end{figure*}

 \begin{figure*}
     \centering
    \includegraphics[width=0.95\linewidth]{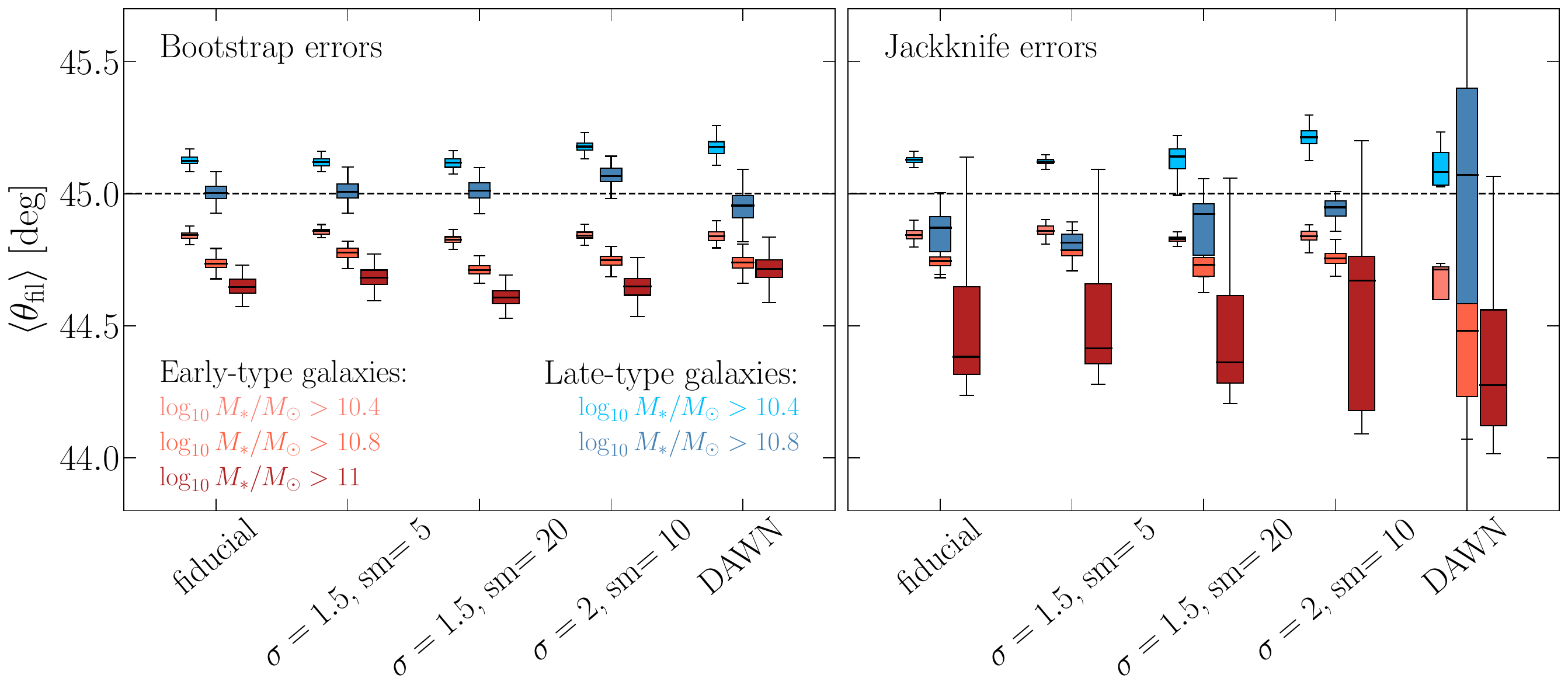}  
     \caption{Box plots displaying the alignment angle $\theta_{\rm fil}$ of early-type (red) and late-type (blue) galaxies at different mass thresholds for various parametrisations of the skeleton extraction (the persistence threshold $\sigma$ and the smoothing length of filaments ${\rm sm}$), with uncertainties estimated from bootstrap (left) or jackknife (right) methods, while keeping fixed the selection of galaxies. The right marker corresponds to the signal measured in the DAWN skeleton (with a persistence $\sigma=1.5$ and filament smoothing length of 10~segments).}
     \label{fig:variations_skeleton}
 \end{figure*}

\begin{figure*}
     \centering
    \includegraphics[width=0.95\linewidth]{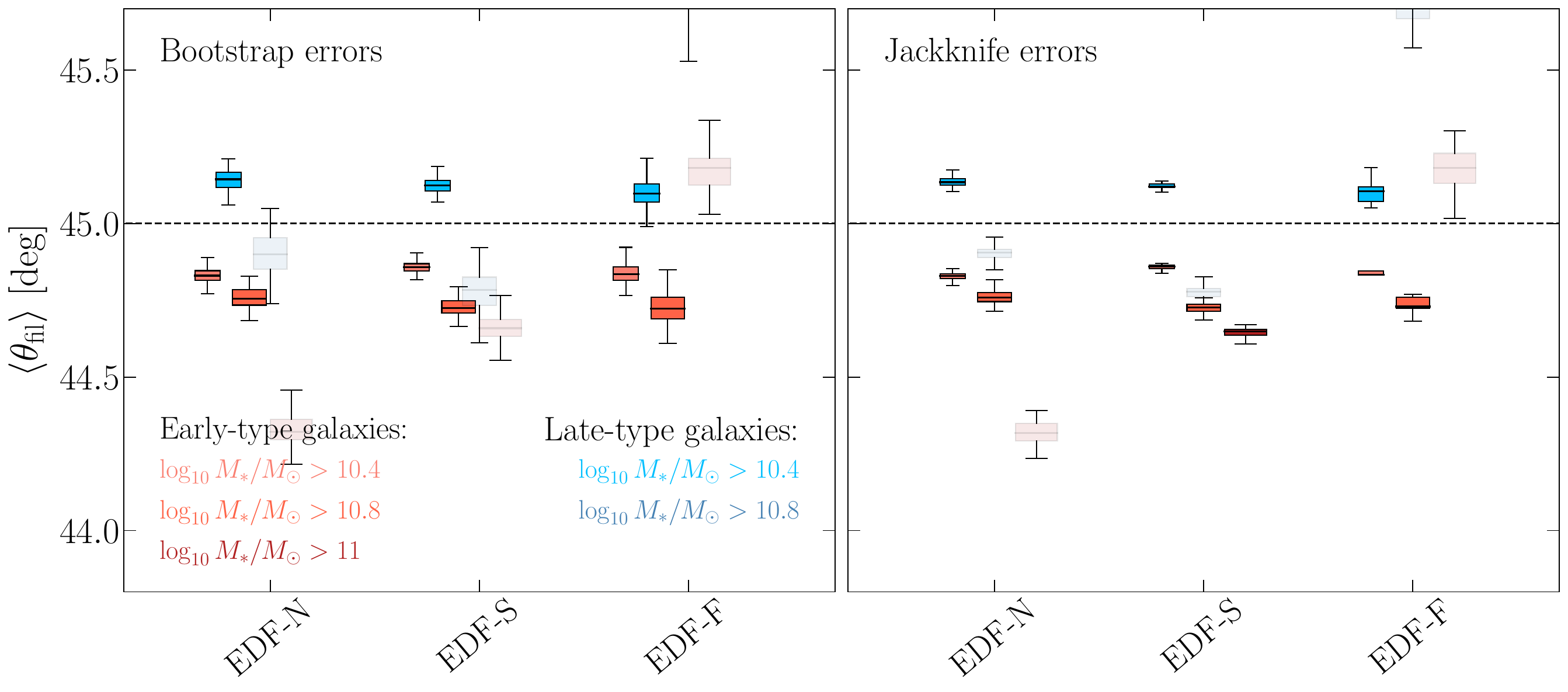}  
     \caption{Box plots displaying the alignment angles $\theta_{\rm fil}$ of early-type (red) and late-type (blue) galaxies at different mass thresholds in the three fields, with uncertainties estimated from bootstrap (left) or jackknife (right) methods. Boxes corresponding to selections with less than 2000~galaxies are shown in transparency. They generally follow the trend but show larger variability.}
     \label{fig:variations_fields}
 \end{figure*}

To assess the robustness of our measurement with respect to field cosmic variance, sample selection and skeleton parametrisation, we recomputed the mean alignment signal at the same mass thresholds as presented in Fig.~\ref{fig:alignment} for various selections and skeleton extractions. 

In Fig.~\ref{fig:variations_selection}, we progressively relax the different criterion on {\tt Sersic\_index}, {\tt Sersic\_ratio} and {\tt Sersic\_angle\_err}, and median distance of galaxies to filaments. The latter has the largest effect on reducing the alignment signal, but overall the signal  remains robust for early-type galaxies, and for late-type galaxies at the mass threshold of $\log_{10}(M_*/M_\odot) > 10.4$.
In addition, we compute both bootstrap uncertainties (drawing 100 samples) and jackknife uncertainties. To do so, we divide each field on the sky in squares of $1~{\rm deg}^2$, and we re-compute the mean angles while excluding one square at a time. In general jackknife and bootstrap statistics are comparable, with two  exceptions: in the highest mass bin for late-type galaxies($\log_{10} {\rm M_{*}/M_\odot}>10.8$) and for early-type galaxies ($\log_{10} {\rm M_{*}/M_\odot}>11$), where the jackknife errors are much larger and make the signal consistent with zero. As a matter of fact, high-mass galaxies are rarer and much more sensitive to sampling variance, which is partly encapsulated in the jackknife estimate. In addition, the high clustering and strong alignment signal of high mass galaxies both make them more sensitive to erroneous cosmic web filament extraction: nearby massive galaxies aligning together with the true underlying dark matter filament could display a strong and systematic mis-alignment if this filament is incorrectly extracted. In turn, this will increase the variance on the sky of the signal. Such variance is not captured by the bootstrap resampling. 

In Fig.~\ref{fig:variations_skeleton} we present the alignment for various flavours of the skeleton extraction. We also display the signal of alignment measured with respect to the skeleton extracted from the DAWN catalogue. To measure it, galaxies in the DAWN and Q1 catalogues are positionally matched, and  galaxies are assigned to the closest tomographic slices based on their redshifts from the DAWN catalogue. We note that the result on the DAWN catalogues are consistent with the Q1 data set or the early-type galaxies, but the error bars are much larger, owing to the smaller area covered by DAWN.

Figure~\ref{fig:variations_fields} presents the alignment signal in the three EDFs independently. The consistency of the signal across fields is an additional guarantee of its robustness. Boxes corresponding to selections with less than 2000~galaxies are shown in transparency. They generally follow the trend but show larger variability.

\begin{figure*}
\centering\includegraphics[width=0.99\textwidth]{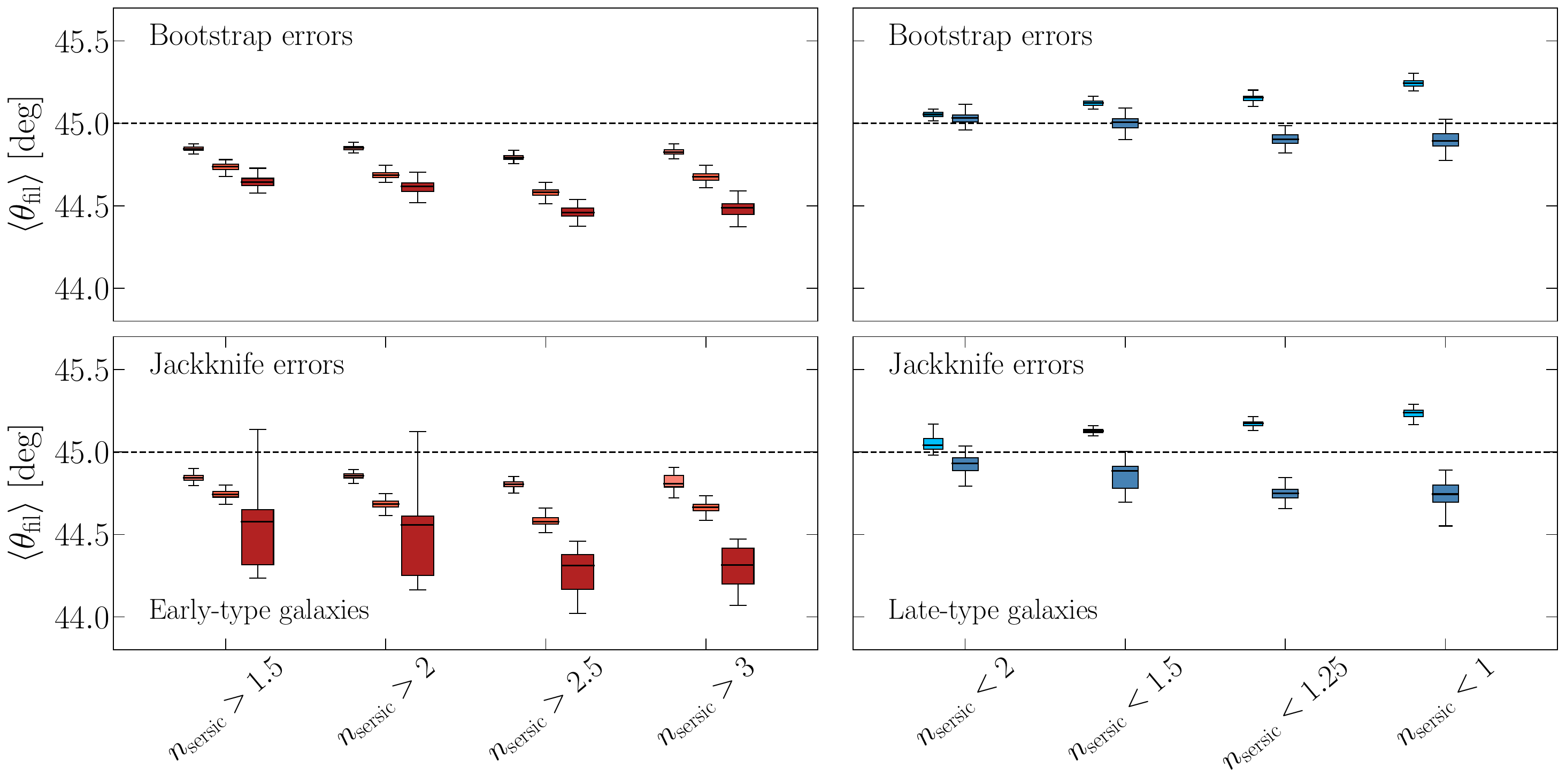}
\caption{Box plots displaying the alignment angles $\theta_{\rm fil}$ of early-type (red) and late-type (blue) galaxies at different mass thresholds based on different selections of early-type and late-type galaxies than chosen in~Fig.~\ref{fig:alignment} (where a threshold $n_{\rm sersic}=1.5$ was used distinguish between both populations).}
\label{fig:alignment_sersicdiff}
\end{figure*}

Finally, Fig.~\ref{fig:alignment_sersicdiff} presents the alignment measurement when selecting early-type and late-type galaxies by varying the cut on the S\'ersic index. In particular, we highlight that the alignment is more significant for galaxies at higher S\'ersic index. For late-type galaxies, the same seems to apply for the alignment of galaxies with $\log_{10} (M_*/M_\odot)>10.4$: a more stringent selection (cutting at $n_{\rm sersic}<1$ instead of $n_{\rm sersic}<1.5$) isolate more galaxies which tend to present the opposite alignment. This is not true for the most massive late-type galaxies ($\log_{10} (M_*/M_\odot)>10.8$).  

 \begin{figure*}
     \centering
    \includegraphics[width=0.95\linewidth]{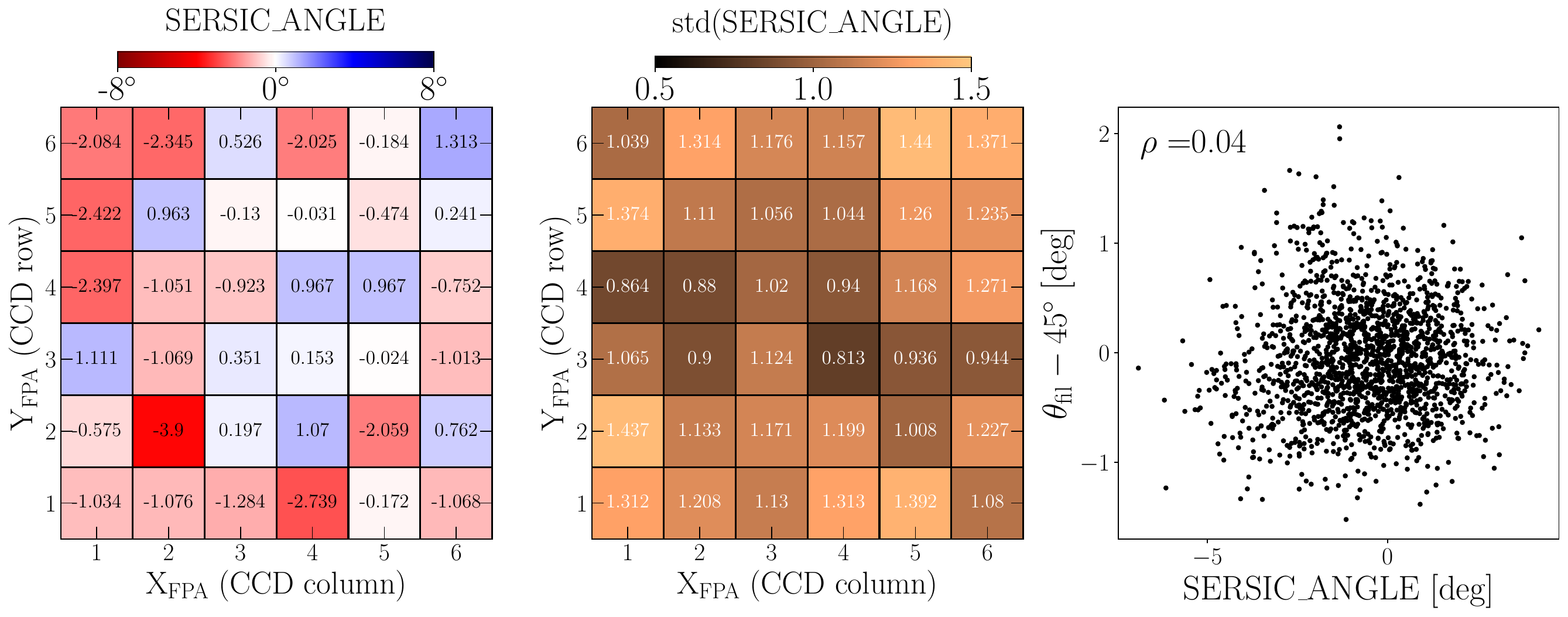}  
     \caption{Variations of the galaxy position angles, as estimated from the S\'ersic fit, as a function of the position on the VIS FPA. The left panel gives the mean value per CCD, and the middle panel the associated standard deviation, both computed from bootstrap resampling of the galaxy population. Although we observe in the mean per CCD some systematic residual deviation from 0, potentially owing to variations of the PSF ellipticity and orientation across the FPA, these deviations are uncorrelated with the mean $\theta_{\rm fil}$ per quadrant, as displayed on the right panel, where each marker corresponds to the measurement in one CCD for one bootstrap realisation. }
     \label{fig:systematic_angle}
 \end{figure*}

\subsection{Variations of position angles across the VIS FPA}
In order to explore possible systematics in the measurement of galaxy position angles, Fig.~\ref{fig:systematic_angle} displays the variation of position angles across the VIS Focal Plane Array (FPA), to investigate if residuals from the VIS processing could be the cause for the detected signal. In particular, we want to investigate systematic variations of the galaxy position angles across the FPA, which could in this case be driven by PSF variations or astrometric distortion residuals. Figure~\ref{fig:systematic_angle} does not show any obvious trend at the scale of the FPA, although the mean galaxy position angle on some VIS Charged Coupled Devices (CCD) depart from~0 by more than~$1\,\sigma$. The right panel shows the absence of correlation between the mean galaxy position angle per CCD and the mean $\theta_{\rm fil}$ per CCD.

\subsection{Standard error on the results presented in Fig.~\ref{fig:2dalign}}

\begin{figure}
    \centering
    \includegraphics[width=0.4\columnwidth]{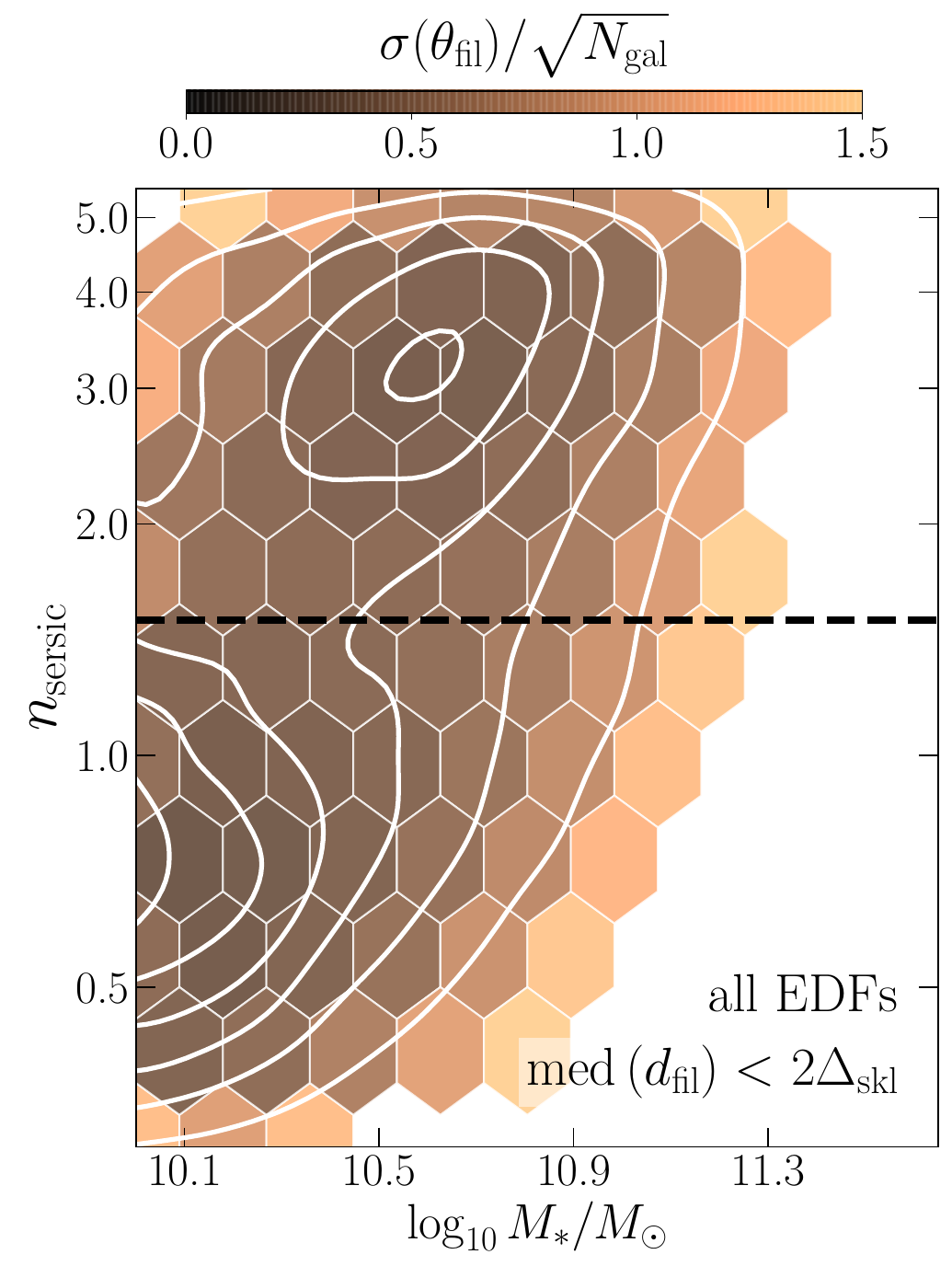}
    \caption{Standard error on the mean angle $\theta_{\rm fil}$ in each hexagon. The mean angle is presented in Fig.~\ref{fig:2dalign}.}
    \label{fig:standerror}
    
\label{LastPage}
\end{figure}

Figure~\ref{fig:standerror} presents the standard error on the mean position angle between galaxy major axes and cosmic web filaments in the plane of S\'ersic index versus mass.

\end{appendix}

\end{document}